\documentclass[12pt]{article}

\input{epsf}
\usepackage[nosort]{cite}
\usepackage[dvips]{graphicx}
\usepackage{amsmath}
\usepackage{epsfig}
\usepackage{amsfonts}
\usepackage{amssymb}
\usepackage{multirow}
\usepackage{array}

\usepackage{epsfig}
\usepackage{amsmath}
\usepackage{amssymb}

\newcommand{\be}{\begin{equation}}
\newcommand{\ee}{\end{equation}}
\newcommand{\bee}{\begin{equation*}}
\newcommand{\eee}{\end{equation*}}
\newcommand{\bea}{\begin{eqnarray}}
\newcommand{\eea}{\end{eqnarray}}
\newcommand{\bean}{\begin{eqnarray*}}
\newcommand{\eean}{\end{eqnarray*}}

\newcommand{\nn}{\nonumber}

\newcommand{\lp}{\left(}
\newcommand{\rp}{\right)}

\begin{document}

\setcounter{page}{0}
\thispagestyle{empty}
%\pagestyle{empty}

%%%%%%%%%%%%%%%%%%%%%%%%%%%%%%%%%%%%%%%%%%%%%%%%%%%%%%%%%%%%%%%%%%%%%%%%%%%%%%%
\begin{flushright}
%astro-ph/yymmnnn
CERN-PH-TH/2010-027
\end{flushright}

\vskip 8pt

\begin{center}
{\bf \LARGE {
Energy Budget of \\
\vskip 8pt
Cosmological First-order Phase Transitions
 }}
\end{center}

\vskip 12pt

\begin{center}
{\bf Jos\'e R. Espinosa$^{a,b}$, Thomas Konstandin$^{b}$,}\\
 \vskip 6pt
{\bf   Jos\'e M. No$^{c}$
  and G\'eraldine  Servant $^{b,c}$ }
\end{center}

\vskip 20pt

\begin{center}

\centerline{$^{a}${\it
ICREA, Instituci\`o Catalana de Recerca i Estudis Avan\c{c}ats,}}
\centerline{\it at IFAE, Universitat Aut{\`o}noma de Barcelona,
08193 Bellaterra, Barcelona, Spain}
\centerline{$^{b}${\it CERN Physics Department, Theory Division, CH-1211 
Geneva 23, Switzerland}}
\centerline{$^{c}${\it Institut de Physique Th\'eorique, CEA/Saclay, F-91191 
Gif-sur-Yvette C\'edex, France}}
\vskip .3cm
\centerline{\tt  jose.espinosa@cern.ch, thomas.markus.konstandin@cern.ch,}
\centerline{\tt jose.miguel.no.redondo@cern.ch, geraldine.servant@cern.ch}
\end{center}

\vskip 13pt

\begin{abstract}
\vskip 3pt
\noindent
The study of the hydrodynamics of bubble growth in first-order phase
transitions is very relevant for electroweak baryogenesis, as the
baryon asymmetry depends sensitively on the bubble wall velocity, and
also for predicting the size of the gravity wave signal resulting from
bubble collisions, which depends on both the bubble wall velocity and
the plasma fluid velocity.  We perform such study in different bubble
expansion regimes, namely deflagrations, detonations, hybrids (steady
states) and runaway solutions (accelerating wall), without relying on
a specific particle physics model. We compute the efficiency of the
transfer of vacuum energy to the bubble wall and the plasma in all
regimes.  We clarify the condition determining the runaway regime and
stress that in most models of strong first-order phase transitions
this will modify expectations for the gravity wave signal. Indeed, in
this case, most of the kinetic energy is concentrated in the wall and
almost no turbulent fluid motions are expected since the surrounding
fluid is kept mostly at rest.
\end{abstract}

\newpage

\tableofcontents

\vskip 13pt

\section{Introduction}

A cosmological first-order phase transition could have far-reaching
consequences and could lead to many interesting phenomena, as for example
electroweak baryogenesis \cite{Cohen:1990py}, primordial magnetic fields
\cite{Hogan:1983zz} or a stochastic background of gravitational waves
\cite{Witten:1984rs,Kosowsky:1991ua,Kamionkowski:1993fg}. All of them are
based on the fact that a first-order phase transition proceeds by bubble
nucleation and, in the process, a large portion of the available vacuum
energy is stored close to the bubble walls. In all these phenomena, some
essential characteristics of the first-order phase transition enter,
namely the velocity of the expanding bubbles and the efficiency
coefficients that keep track of the energy budget of the transition, that 
is, how the free energy of the Higgs field $\phi$ initially available is 
distributed among  bulk fluid motion, thermal energy of the plasma 
and gradient/kinetic energy of the Higgs field.

Quite generally, treatments of the bubble wall velocity assume that
the expansion of the bubbles is hindered by some form of friction,
such that the bubble wall reaches a constant speed after a rather
short time of order $\sim 1/m$, where $m$ represents the typical mass
scale associated with the transition (i.e. electroweak mass scale in
the case of the electroweak phase transition). Assuming that the free
energy of the Higgs field is released into the plasma, a hydrodynamic
treatment of the plasma can be used to determine the fluid motion
\cite{Landau, Steinhardt:1981ct}.  However, as long as the microscopic
mechanism of friction is unknown, this approach leaves one free
parameter, typically the wall velocity. In \cite{Steinhardt:1981ct} it
was argued that the velocity should be fixed by the Chapman-Jouguet
condition, like in chemical combustion, and this condition was
subsequently used in many studies of gravitational wave production
\cite{Apreda:2001us}. The Chapman-Jouguet condition generally leads to
supersonic wall velocities with a rarefaction wave behind the bubble
wall. This scenario favors gravitational wave production, since fast
moving walls are essential for the production of gravitational
radiation in bubble collisions~\cite{Kosowsky:1991ua,
Kamionkowski:1993fg, Caprini:2007xq, Huber:2008hg, Caprini:2009fx },
turbulence~\cite{Kosowsky:2001xp,Caprini:2006jb} or magnetic
fields~\cite{Caprini:2006jb}.

Nevertheless, while being observed in chemical combustion, the
Chapman-Jouguet condition is unrealistic for cosmological phase
transitions \cite{Laine:1993ey}. To replace the Chapman-Jouguet condition,
friction has been studied through a phenomenological parametrization in the Higgs
equation of motion \cite{Moore:1995si, Ignatius:1993qn}. 
Extensive simulations of hydrodynamic equations in
conjunction with an equation for the Higgs field have been used to
identify the stable expansion modes of the bubbles~\cite{Laine:1993ey,
Ignatius:1993qn, KurkiSuonio:1984ba, Gyulassy:1983rq, KurkiSuonio:1995pp}.
It turns out that
many solutions which are viable close to the phase transition front, where
the free energy of the Higgs field is released into the plasma, are not
stable globally. 

The friction coefficient appearing in the Higgs equation can be
determined by solving Boltzmann-type equations close to the phase
transition front~\cite{Moore:1995si, John:2000zq}. This approach leads to
subsonic wall velocities in the Standard Model (SM) and its supersymmetric
extension (MSSM) and is widely used in studies of MSSM electroweak
baryogenesis~\cite{Cline:2000nw, Carena:2000id, Carena:2002ss,
Konstandin:2005cd, Cirigliano:2006dg}. Subsonic wall velocities are
crucial in electroweak baryogenesis, since this mechanism is based on
diffusion of particle asymmetries into the plasma in front of the bubble
wall and for a too fast wall, there is no time to build up a
baryon asymmetry.

In strong first-order phase transitions, the friction exerted by the
plasma on the wall might not be sufficient to prevent the bubble wall
from a runaway behavior in which the wall keeps accelerating, toward
ultra-relativistic velocities, as pointed out recently in \cite{BM}. We
also study this runaway regime and discuss the energy balance in that
case.

Our general goal in this paper is to present a unified picture of all
the different regimes. We go beyond the (unjustified) Chapman-Jouguet
assumption and provide general formula for the wall velocity, the
fluid velocity and the efficiency factors accounting for the
distribution of energy among bulk fluid motion, thermal energy of the
plasma and gradient/kinetic energy of the Higgs field.  The paper is
organized as follows. We review the hydrodynamic treatment of the
plasma in secs.~\ref{sec_hydro} and \ref{sec_detdef}, and subsequently
obtain the efficiency coefficients in sec. \ref{sec_eff}. In
sec.~\ref{sec_static} we present a simplified treatment of the Higgs
equation (a similar one has been carried out recently in
\cite{Megevand:2009ut}). We present detailed results both for the wall
velocity and the efficiency coefficients that are required for the
determination of gravitational wave spectra.  We discuss the runaway
regime in sec.~\ref{sec_runaway} and summarize our results on the energy
budget of first-order phase transitions in sec.~\ref{sec:summary}. We
conclude in sec.~\ref{sec:conclusion} and provide numerical fits to the 
efficiency coefficients in Appendix~A.

\section{Hydrodynamic relations\label{sec_hydro}}

In this section we introduce the basic concepts and set up the notation
used for the hydrodynamic analysis of the combined ``wall-plasma"
system \cite{Landau, Steinhardt:1981ct, Ignatius:1993qn}.

\subsection{Basic concepts \label{subsec_basic}}

The energy-momentum
tensor of the Higgs field $\phi $ is given by
\be
\label{eq:TmunuHiggs}
T_{\mu\nu}^{\phi} = \partial_{\mu}\phi \partial_{\nu}\phi -g_{\mu\nu} 
\left[ \frac{1}{2} \partial_{\rho}\phi \partial^{\rho}\phi - V_0 (\phi)\right],
\ee
where $V_0 (\phi) $ is the renormalized vacuum potential. The energy
momentum-tensor of the plasma is given by
\be
 T_{\mu\nu}^{plasma} = \sum_{i} \int \frac{d^3 k}{(2\pi)^3 E_{i}} 
k_\mu k_\nu f_{i}(k,x), 
\ee
where the sum is carried out over the species in the plasma and
$f_{i}(k,x) $ is the distribution function for each species. If the
plasma is locally in equilibrium (perfect fluid) this can be
parametrized as
\be
\label{eq:Tmunu}
T_{\mu\nu}^{plasma} = w \, u_\mu u_\nu  - g_{\mu\nu} \, p, 
\ee
where $w$ and $p$ are  the plasma enthalpy and pressure, respectively.
The quantity $u_\mu$ is the four-velocity field of the plasma, related to 
the
three-velocity $\mathbf{v}$ by
\be
u_\mu = \frac{(1, \mathbf{v} )}{\sqrt{1-\mathbf{v}^2}}
= (\gamma, \gamma\mathbf{v} ) \ .
\ee
A constant $\phi$ background contributes to the total pressure [see
eq.~(1)] and from now on we will use $p$ for this total pressure,
including such contribution.
 
The enthalpy $w$, the entropy density $\sigma$ and the energy density
$e$ are defined by
\be
w \equiv T\frac{\partial p}{\partial T}\ ,\quad
\sigma \equiv \frac{\partial p}{\partial T}\ ,\quad
e\equiv T\frac{\partial p}{\partial T} -p\ ,
\ee
where $T$ is the plasma temperature. One then has 
\be
w = e + p\ .
\ee
Conservation of energy-momentum is given by 
\be
\label{eq:T_cons}
\partial^\mu T_{\mu\nu} = \partial^\mu T_{\mu\nu}^{\phi} 
+ \partial^\mu T_{\mu\nu}^{\mathrm{plasma}} = 0 \ .
\ee
We are primarily interested in a system with a constant wall
velocity and, assuming there is no time-dependence,
eq.~(\ref{eq:T_cons}) reads in the wall frame (with the wall and fluid
velocities aligned in the $z$ direction)
\be
\partial_z T^{zz} = \partial_z T^{z0} = 0\ .
\ee
Integrating these equations across the phase transition front and denoting 
the phases in front and behind the wall by
subscripts $+$ (symmetric phase) and $-$ (broken phase) one gets
the matching equations (in the wall frame):
\be
\label{eq:wall_constr}
w_+ v^2_+ \gamma^2_+ + p_+  = w_- v^2_- \gamma^2_- + p_-\ ,
\quad 
w_+ v_+ \gamma^2_+ = w_- v_- \gamma^2_-\ .  
\ee
From these equations we can obtain the relations
\be
\label{eq:vvs0}
v_+ v_- = \frac{p_+  - p_-}{e_+ - e_-}\ , \quad
\frac{v_+}{ v_-} = \frac{e_- + p_+ }{e_+ + p_-}\ . 
\ee
To proceed further, one needs to assume a specific equation of state
(EoS) for the plasma. 

\subsection{Equation of state \label{subsec_EoS}}

Usually the plasma is well
described by a relativistic gas approximation. In the symmetric phase, 
\be
\label{eosm}
p_+ = \frac{1}{3}a_+ T_+^4 - \epsilon\ ,\quad
e_+ = a_+ T_+^4 +\epsilon\ ,
\ee
where $\epsilon$ denotes the
false-vacuum energy resulting from the Higgs potential (defined to be
zero in the broken, $T=0$, true-minimum phase). 
While in the broken phase
\be
\label{eosp}
p_- = \frac{1}{3}a_- T_-^4\ , \quad
e_- = a_- T_-^4 \ ,
\ee
with a different number of light degrees of freedom across the wall and 
therefore different values $a_+$ and $a_-$ (with $a_+>a_-$) 
and different temperatures on 
both sides of the wall. These expressions correspond to the 
so-called bag equation of state. 

In the general case, the free energy (${\cal F}=-p$, also called sometimes 
effective potential, including finite temperature corrections) of a 
plasma of particles with arbitrary masses $m_i(\phi)$ is given, in the 
non-interacting-gas approximation, by
\bea
{\cal F} &=& V_0+\int \frac{d^3p}{(2\pi)^3}\sum_i N_i\log\left[
1\mp e^{-E_i/T}\right]\nn\\
&=& V_0+\frac{T^4}{2 \pi^2} \sum_i N_i \, Y_{b/f}(m_i/T)\ ,  
\eea
where $V_0$ is the $T=0$ effective potential, $E_i^2=p^2+m_i^2$,
\be
Y_{b/f}(x) = \int_0^\infty dy \, y^2 \log [1 \mp \exp(-\sqrt{x^2 + y^2})],
\ee
the $-/+$ signs in the last equation hold for bosons/fermions, the
index $i$ denotes the different species and $N_i$ the internal degrees of
freedom (that in this notation is negative for fermions). 
Species that are
light compared to $T$ behave as a relativistic gas while species much
heavier than $T$ are Boltzmann suppressed in the plasma and can be
neglected. As is well known, for small $m_i/T$ the function $Y_{b/f}$ 
tends to a constant ($-\pi^4/45$ for bosons and $7\pi^4/360$ for fermions) 
so that $a_\pm$ in the bag EoS are explicitly given by 
\be
a_{\pm}=\frac{\pi^2}{30}\sum_{{\mathrm{light}}~ 
i}\left[N^b_i+\frac{7}{8}|N^f_i|\right]\ ; 
\ee
while for large $m_i/T$ one has $Y_{b/f}(m_i/T)\sim \pm \exp{(-m_i/T)}$.
Hence, only
particles that have masses comparable to $T$ can cause deviations from the
bag EoS. In many cases this deviation is small because the free-energy is
dominated by a large number of light degrees of freedom. For example in
the Standard Model, the $W^\pm$ and $Z^0$ bosons contribute to this
deviation from the bag EoS (and to a lesser extent the tops) but
altogether the modification in the energy density is at the percent level
and hence small.

If the deviations from the bag EoS are not so small one can still
parametrize the plasma behavior in terms of quantities that mimic the
bag EoS ones. In particular, from the free-energy (${\cal F}=-p$) we
can define 
\be
\label{aepsgen}
a_\pm\equiv 
\frac{3}{4T_\pm^3}\left.\frac{\partial p}{\partial 
T}\right|_\pm =
\frac{3\omega_\pm}{4T_\pm^4}
\ ,\quad
\epsilon_\pm\equiv \frac{1}{4}(e_\pm -3p_\pm)\ .
\ee
In the special case of bag EoS, these definitions reproduce the correct 
$a_\pm$ and give $\epsilon_+=\epsilon$ and $\epsilon_-=0$. 
In terms of these quantities one can write
\be
\label{eosg}
p_\pm = \frac{1}{3}a_\pm T_\pm^4 - \epsilon_\pm\ ,\quad
e_\pm = a_\pm T_\pm^4 +\epsilon_\pm\ ,
\ee
but now $a_\pm$ and $\epsilon_\pm$ are $T$-dependent quantities
and should be interpreted with some care. By their definition 
(\ref{aepsgen}) they  satisfy:
\be
\frac{\partial \epsilon_\pm}{\partial T_\pm}=\frac{T^4_\pm}{3}
\frac{\partial a_\pm}{\partial T_\pm}\ .
\ee
There is a certain arbitrariness in the definition of $a_\pm$ but ours
is especially convenient because some formulas we discuss in
sect.~\ref{sec_static} can be easily generalized even if the bag EoS
is not valid.

Using the bag equations of state (\ref{eosm}) and (\ref{eosp}) in 
eq.~(\ref{eq:vvs0}) we get
\bea
\label{eq:vvs}
v_+ v_- &=& \frac{1 - (1-3\alpha_+) r }
{3 - 3( 1 + \alpha_+ )r}, \nn \\ 
\frac{v_+}{ v_-} &=& \frac{3  + (1-3\alpha_+) r}
{1 + 3(1 + \alpha_+ )r}, 
\eea
where we defined 
\be
\alpha_+\equiv\frac{\epsilon}{a_+ T_+^4}\ ,\quad  
r\equiv\frac{a_+T_+^4}{a_-T_-^4}\ .
\ee 
The quantity $\alpha_+$ is the ratio of the vacuum energy to the radiation 
energy density and typically characterizes the ``strength" of the phase 
transition: the larger $\alpha_+$ the stronger the phase transition.
These two equations can be combined to give
\be
\label{eq:vvs2}
v_+ = \frac{1}{1+\alpha_+}\left[ \left(\frac{v_-}{2}+\frac{1}{6
v_-}\right) \pm \sqrt{\left(\frac{v_-}{2}+\frac{1}{6 v_-}\right)^2 +
\alpha_+^2 +\frac{2}{3}\alpha_+ - \frac{1}{3}} \right],
\ee
so that there are two branches of solutions, corresponding to the
$\pm $ signs in eq. (\ref{eq:vvs2}). In the general case, with some 
deviation from the bag EoS,
eqs.~(\ref{eq:vvs}) and (\ref{eq:vvs2}) still apply, with $r$ defined as
before and $\alpha_+\equiv (\epsilon_+-\epsilon_-)/(a_+ T_+^4)$. 
%The behaviour of the velocity solutions as shown by
%Figure~\ref{fig:vpvm_contours} will therefore be the same.

\begin{figure}[ht]
\begin{center}
\includegraphics[width=0.65\textwidth, clip ]{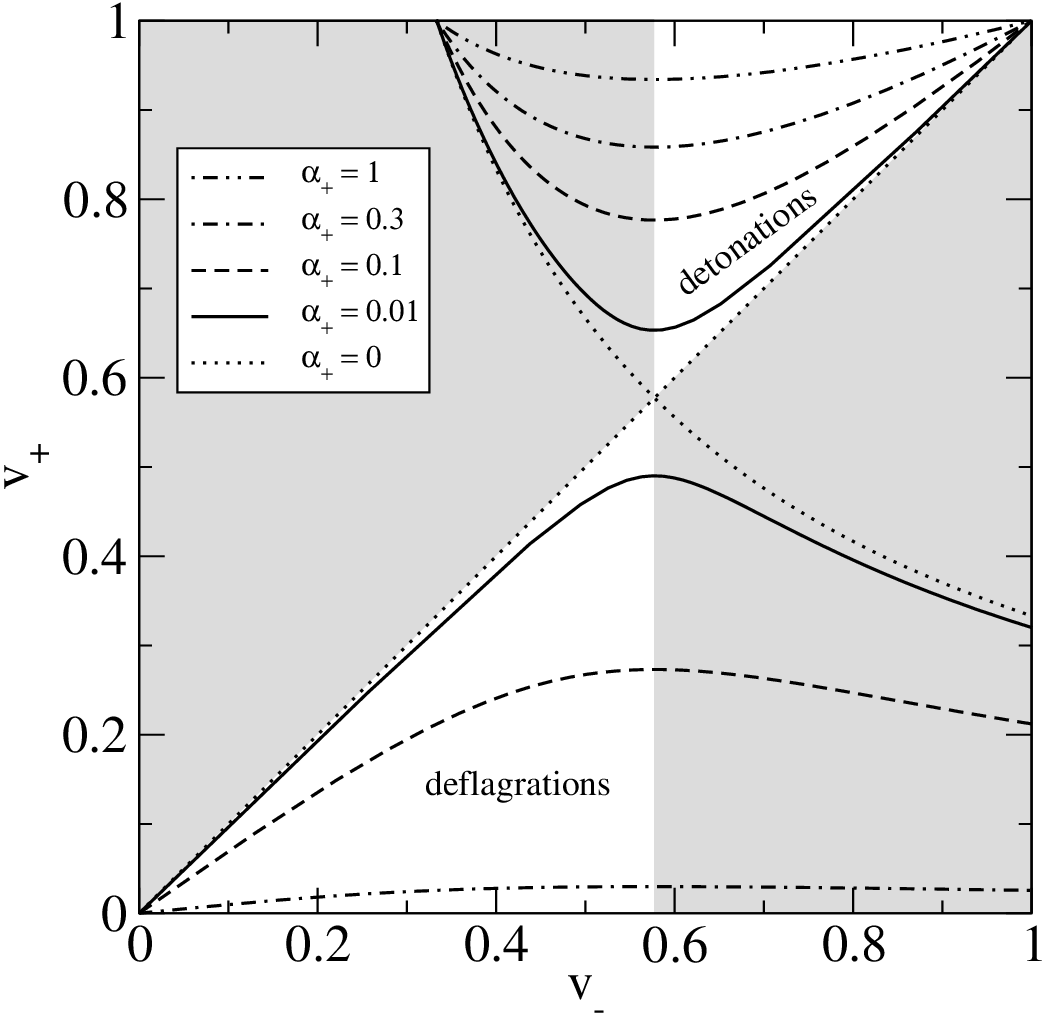}
\caption{\label{fig:vpvm_contours}
\small Contours of the fluid velocities $v_+$ and 
$v_-$ in the wall frame for fixed $\alpha_+$. In
the shaded region in the top-left no consistent solutions to
hydrodynamic equations exist. Flow profiles in the shaded region in the
bottom-right decay into hybrid solutions, with $v_-=c_s^-$ (see 
sect.~\ref{subsec:hybrids}).}
\end{center}
\end{figure}

Figure~\ref{fig:vpvm_contours} shows these solutions for several values
of $\alpha_+$.  The first branch
(with positive sign)  gives solutions of detonation type, $v_+ > v_-$, 
corresponding to $r<\frac{1}{1+3\alpha_+}$.  
The function $v_+$ has a minimum at $v_-=c_s$. Notice that the point
$v_+ = v_- = 1$ is always a solution of this kind, with $r=\frac{1}{1 +
3\alpha_+}$; besides, in the limit $r \to 0$ one obtains $v_+=1$ and
$v_-=1/3$.  The second branch of solutions exists only if $\alpha_+ <
1/3$, and is of deflagration type, $v_+ < v_-$ (it ranges over all values
of $v_-$). Also for deflagrations, the function $v_+$ has an extremum at
$v_-=c_s$; the endpoint, $v_-=1$, is given by the limit $r \to \infty$ and
yields $v_+=\frac{1-3\alpha_+}{3+3\alpha_+}$. The limiting case
$\alpha_+=0$ corresponds to $v_+ v_-=1/3$, connects both regions and will
be relevant for the discussion of shock fronts in the next section.

To determine all relevant quantities in terms of the phase transition
parameters, one still has to determine one of the two velocities, or
equivalently, relate $T_+$ to $T_-$; this will be the subject of sec.
\ref{sec_static}. Before doing so, we want to provide a more detailed
description and a physical understanding of the above solutions as well as
of the hybrid regime. Hence, for now, we treat the wall velocity as a free
parameter, without any further assumptions about the microscopic
properties of the plasma. In sections \ref{sec_detdef} and \ref{sec_eff}
we present the analysis of the different types of solutions for the motion
of the plasma fluid. We will be particularly interested in the efficiency
factor $\kappa$ (that measures how much of the vacuum energy goes into
bulk kinetic energy and is a crucial quantity to determine gravitational
wave production) as a function of the wall velocity. To define $\kappa$,
we first need the relativistic fluid equations for the plasma that we
derive now.

\subsection{Relativistic fluid equations and bulk kinetic 
energy\label{subsec_kappa}}

The starting point are the continuity equations (\ref{eq:T_cons}) that, 
for a general plasma of form (\ref{eq:Tmunu}), read (see e.g.~\cite{Landau})
\be
\label{eq:T_cons2}
\partial_\mu T^{\mu\nu} = u^\nu \partial_\mu (u^\mu \, w)
+ u^\mu \, w \partial_\mu u^\nu - \partial^\nu \, p.
\ee
Projecting along the flow then leads, using $u_\mu \partial_\nu
u^\mu = 0$, to
\be
\label{eq:Eflux_cons}
\partial_\mu (u^\mu \, w) - u_\mu \partial^\mu \, p = 0.
\ee

Consider next the projection perpendicular to the flow with some
space-like vector $\bar u = \gamma (v, \mathbf{v}/v)$ such that
$\bar u_\mu u^\mu = 0$, $\bar u^2=-1$. This turns (\ref{eq:T_cons2}) into 
the relativistic Euler equation
\be
\label{eq:Euler}
\bar u^\nu u^\mu \, w \partial_\mu u_\nu - \bar u^\nu \partial_\nu \, p=0.
\ee
In the following we assume a spherically symmetric configuration.
Additionally, because there is no characteristic distance scale in the
problem, the solution should be a similarity solution that depends only on
the combination $\xi = r/t$ where $r$ is the distance from the center of
the bubble and $t$ is the time since nucleation.\footnote{
The matching equations (\ref{eq:wall_constr}) can also be derived using 
energy-momentum conservation for similarity solutions in the rest frame.}
$\xi $ is thus the
velocity of a given point in the wave profile and the particles at the
point described by $\xi$ in the wave  profile move with velocity $v(\xi)$, 
which is therefore the fluid velocity in the frame of the bubble center.
This turns the different gradients into 
\be
 u_\mu \partial^\mu = - \frac{\gamma}{t} (\xi - v) \partial_\xi, \quad
\bar u_\mu \partial^\mu = \frac{\gamma}{t} (1 - \xi \, v) \partial_\xi.
\ee
Finally, this yields for eqs.~(\ref{eq:Eflux_cons}) and
(\ref{eq:Euler})
\bea
\label{eq:Eulerx}
(\xi - v ) \frac{\partial_\xi e}{w} &=& 
2\frac{v}{\xi} + 
[1 - \gamma^2 v (\xi - v)]\partial_\xi v\ , \nn \\
(1 - v \xi ) \frac{\partial_\xi p}{w} &=& 
\gamma^2 (\xi - v ) \partial_\xi v.
\eea
The derivatives $\partial_\xi e$ and $\partial_\xi p$ can be related 
through the speed of sound in the plasma, $c_s^2\equiv (dp/dT)/(de/dT)$,
so as to get the central equation describing the velocity profile:
\be
\label{eq:flow}
2\frac{v}{\xi} = \gamma^2 (1- v  \xi) 
\left[ \frac{\mu^2}{c_s^2} - 1 \right] \partial_\xi v,  
\ee
with $\mu$ the Lorentz-transformed  fluid velocity
\be
\mu(\xi, v)=\frac{\xi-v}{1- \xi v}\ .
\label{lorentz}
\ee 
In general, $c_s^2$ depends on the EoS for the plasma, being 
$c_s^2=1/3$ in the bag case. In the general case, $c_s^2$ will be 
$\xi$-dependent, although in many cases of interest deviations from $1/3$
will be small.

Eq.~(\ref{eq:flow}) can then be solved (with the appropriate 
boundary conditions) to yield the velocity profile $v(\xi)$ of the plasma. 
Subsequently, eqs.~(\ref{eq:Eulerx}) can be integrated to yield
\be
\label{eq:enthalpy}
w(\xi) = w_0 \, \exp \left[  
\int_{v_0}^{v(\xi)}\left(1+\frac{1}{c_s^2}\right) \gamma^2 \, \mu \, dv 
\right].
\ee

In the calculation of the gravitational radiation produced in the phase 
transition one needs to compute the kinetic energy in the bulk motion of 
the plasma. We have now all ingredients necessary to perform such 
calculation. The ratio of
that bulk kinetic energy over the vacuum energy gives the efficiency
factor $\kappa$ as 
\be
\label{eq:defKap}
\kappa = \frac{3}{\epsilon \xi_w^3} 
\int w (\xi) v^2 \gamma^2 \, \xi^2 \, d\xi\ ,
\ee
where $\xi_w$ is the velocity of the bubble wall. Notice that this 
definition coincides with the expression used in the gravitational wave 
literature, that is given by $\kappa = \frac{3}{\epsilon R_w^3} \int w\, 
v^2 \gamma^2\,R^2 dR$, but differs from the definition used in
ref.~\cite{Kamionkowski:1993fg} by a factor $\xi_w^3$. 

We also numerically check  energy conservation: Integration of $T_{00}$
over a region larger than the bubble (including the shock front) is
constant in time, giving
\be
\int\left[ ( \gamma^2 - \frac{1}{4} )  w  - \frac{3}{4} w_N \right]   
\xi^2 d\xi =  \frac\epsilon3 \xi_w^3,
\ee
where $w_N$ denotes the enthalpy at nucleation temperature far in
front of the wall. This implies that the energy which is not
transformed into kinetic bulk motion, but is used instead to increase the 
thermal energy, is
\be
1 - \kappa = \frac{3}{\epsilon \xi_w^3} 
\int \frac34 ( w   - w_N )   \xi^2 d\xi =
\frac{3}{\epsilon \xi_w^3} 
\int ( e   - e_N )   \xi^2 d\xi.
\ee

\section{Detonations, deflagrations and hybrids\label{sec_detdef}}

We can now use the previous fluid equations to describe the different
kinds of solutions for the motion of the plasma disturbed by the moving
phase transition wall. In the discussion below, the sound velocity in the
plasma plays a very relevant role. This velocity will in general depend on
$\xi$ and it is convenient to distinguish its asymptotic values in
the symmetric and broken phases. We denote those two velocities by
$c_s^\pm$. In many cases, we expect the bag EoS to hold in the symmetric
phase and therefore $c_s^+=1/\sqrt{3}$.

\begin{figure}[ht]
\begin{center}
\includegraphics[width=0.8\textwidth, clip ]{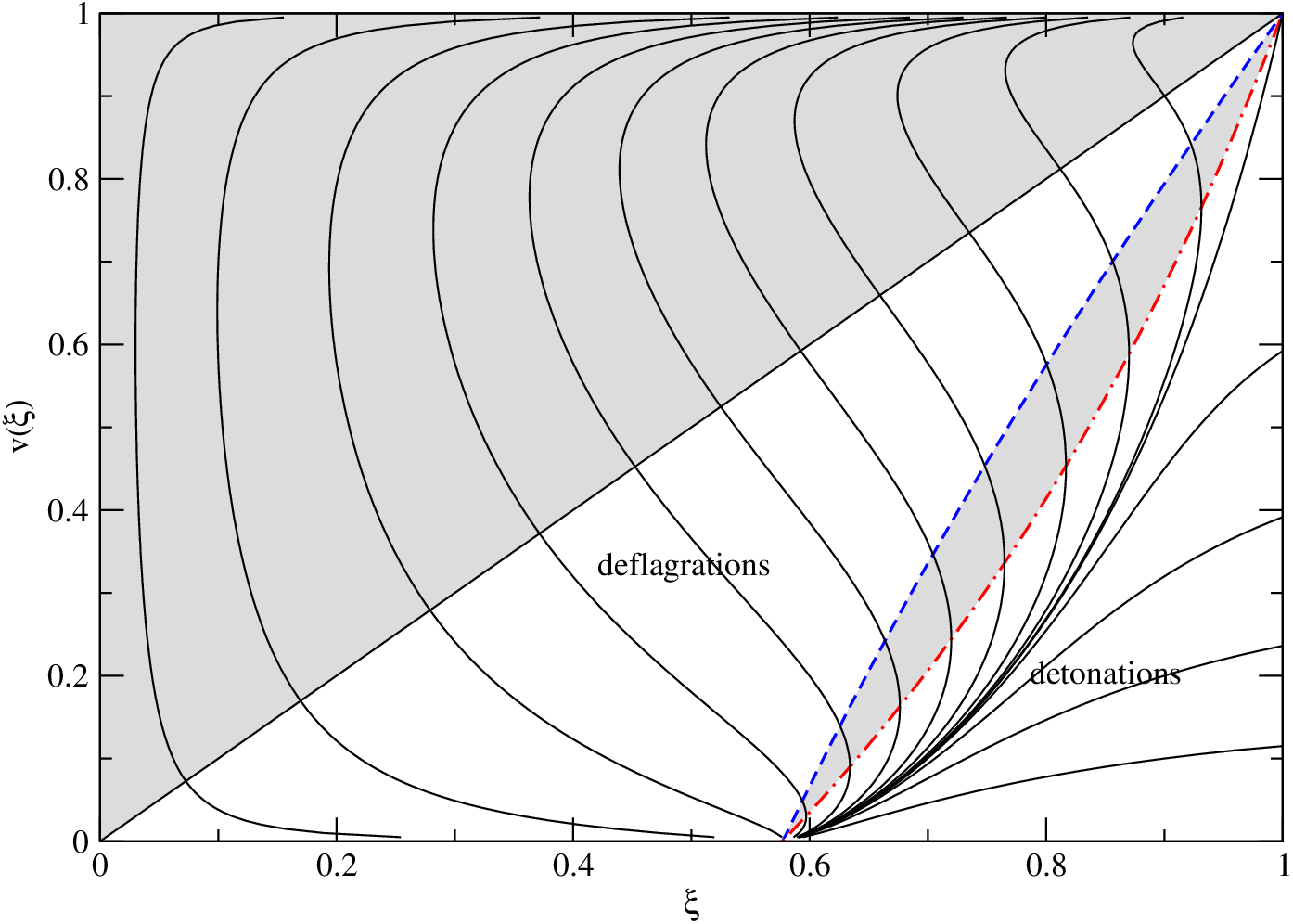}
\caption{\label{fig:prof}
\small General profiles of the fluid velocity $v(\xi)$ in the frame of 
the bubble center (with $c_s^2=1/3$). 
Detonation curves start below $\mu(\xi,v)=c_s$ (dashed-doted curve) and 
end at $(\xi,v)=(c_s,0)$. Deflagration curves start below $v=\xi$ and end 
at $\mu(\xi,v) \xi = c_s^2$ (dashed curve) corresponding to the shock 
front, as explained in the text. There are no consistent solutions in the 
shaded regions.}
\end{center}
\end{figure}

Before embarking in the discussion of the different types of velocity
profiles, it proves useful to study first in more detail the profile
eq.~(\ref{eq:flow}) without worrying about physical boundary conditions.
The different curves in Fig.~\ref{fig:prof} are obtained by solving for
$\xi$ as a function of $v$ [instead of the more physically meaningful
$v(\xi)$, the plasma velocity profile] using arbitrary boundary conditions
and setting $c_s=1/\sqrt{3}$. This procedure has the advantage that
$\xi(v)$ is a single-valued function. The meaning of the different regions
will be explained later on. We see two different fixed points: one for
$\xi=c_s$ and $v=0$ and the other for $\xi=v=1$. This structure of fixed
points can be understood analytically in a simple way by introducing an
auxiliary quantity $\tau$ to describe parametrically these curves in the
plane $(\xi,v)$. Using such parameter, eq.~(\ref{eq:flow}) can be split in
the two simpler equations
\bea
\frac{d v}{d \tau} & = & 2 v c_s^2 (1-v^2) (1-\xi v)\ , \nonumber\\
\frac{d \xi}{d \tau} & = & \xi [(\xi-v)^2-c_s^2(1-\xi v)^2]\ ,
\eea
which clearly display the mentioned fixed points. The point $(c_s,0)$ is 
in fact what is technically called an improper node and all curves 
approach it tangentially to the $v=0$ line. One also sees that 
$dv/d\tau>0$ so that along each curve $v$ grows monotonically with $\tau$.
In the more general case with a $\xi$-dependent $c_s$ deviating somewhat 
from $1/\sqrt{3}$ one expects quantitative changes in these curves but the 
same qualitative behavior [with the fixed point at $(c_s^-,0)$].

Physical velocity profiles for expanding bubbles have to go to zero at 
some distance in front and behind the bubble wall and, in view of 
Fig.~\ref{fig:prof}, this clearly requires some discontinuous jump, which 
usually can take place right at the phase transition front, where many 
other quantities also jump. The different types of possible velocity 
profiles that result are discussed next. 

\subsection{Detonations}

\begin{figure}[ht]
\begin{center}
\includegraphics[width=0.8\textwidth, clip ]{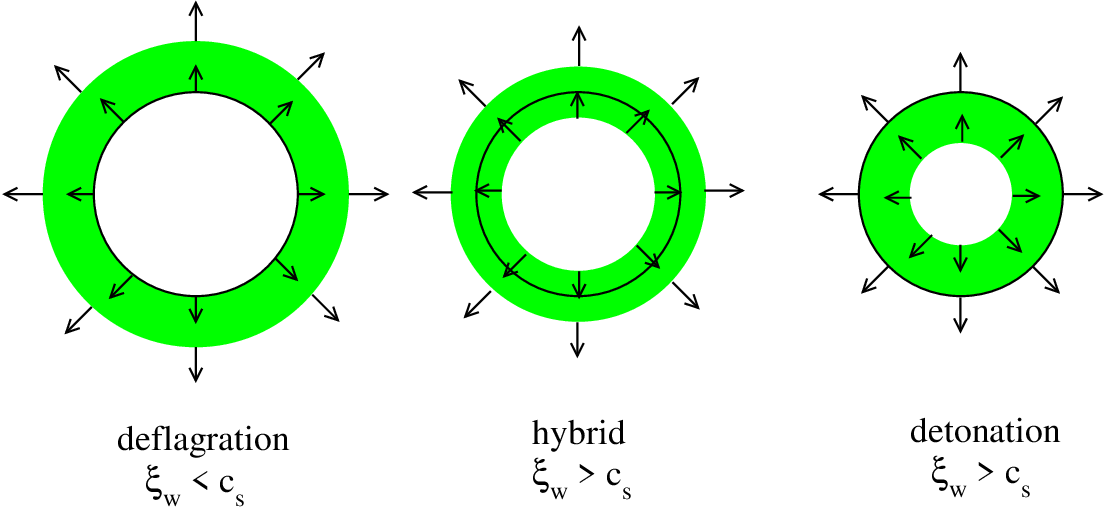}
\caption{\label{fig:bubbles}
\small Pictorial representation of expanding bubbles of different types. 
The black circle is the phase interface (bubble wall). In 
green we show the region of non-zero fluid velocity.}
\end{center}
\end{figure}

\begin{figure}[ht!]
\begin{center}
\includegraphics[width=0.925\textwidth,clip]{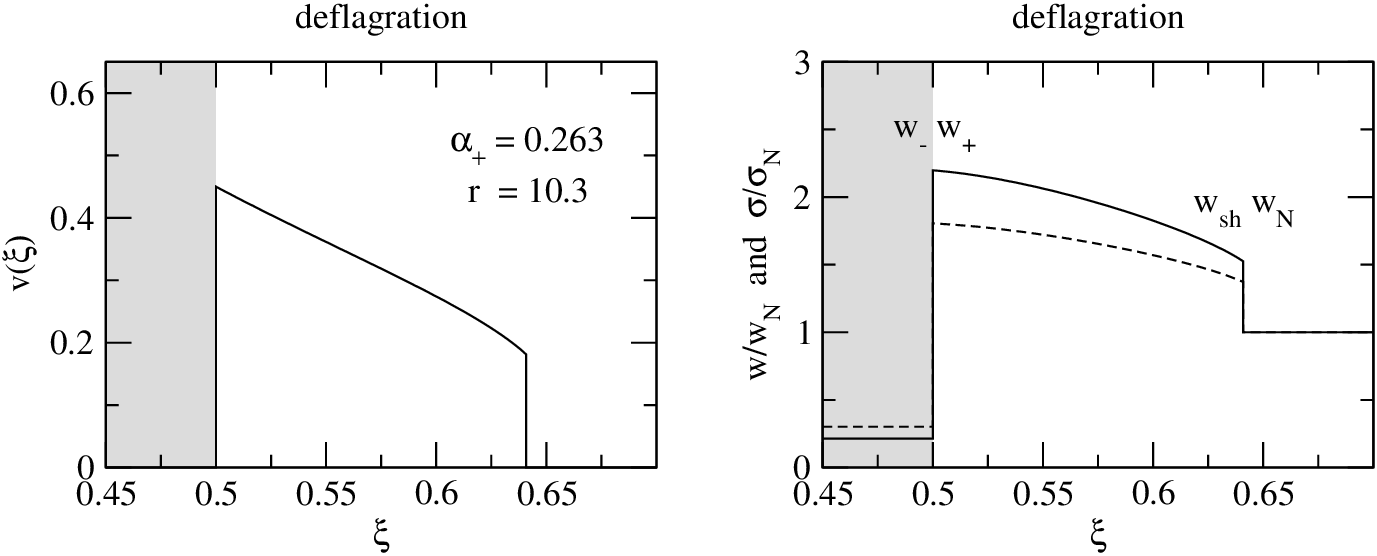}
\vskip 0.3 cm
\includegraphics[width=0.925\textwidth,clip]{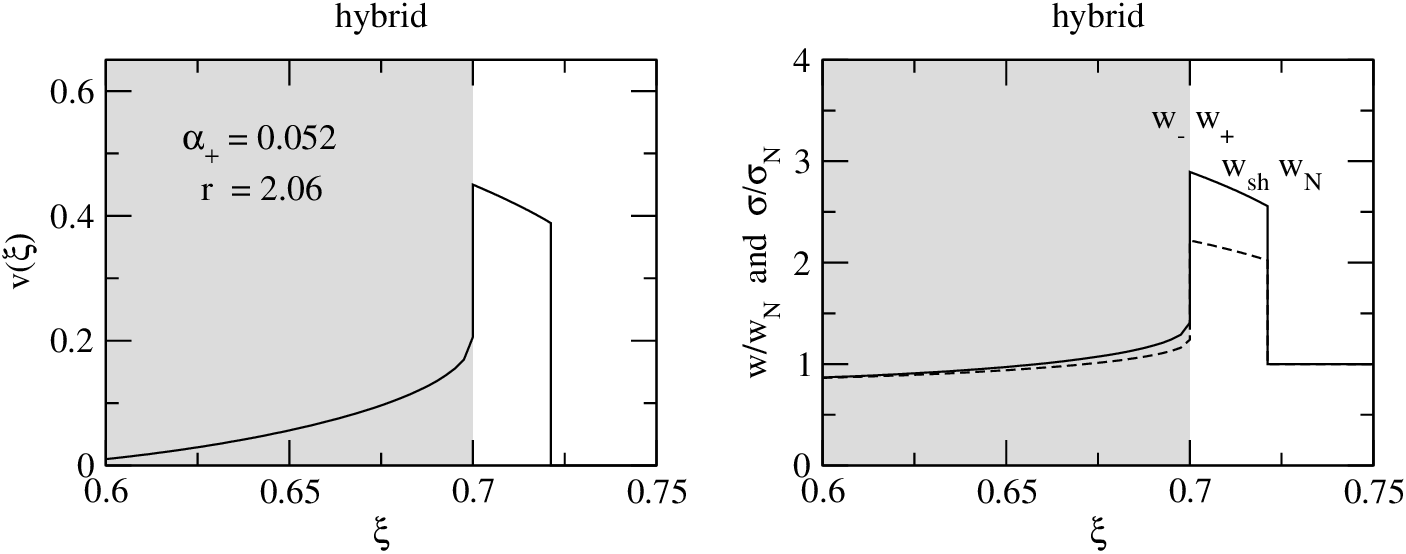}
\vskip 0.3 cm
\includegraphics[width=0.925\textwidth,clip]{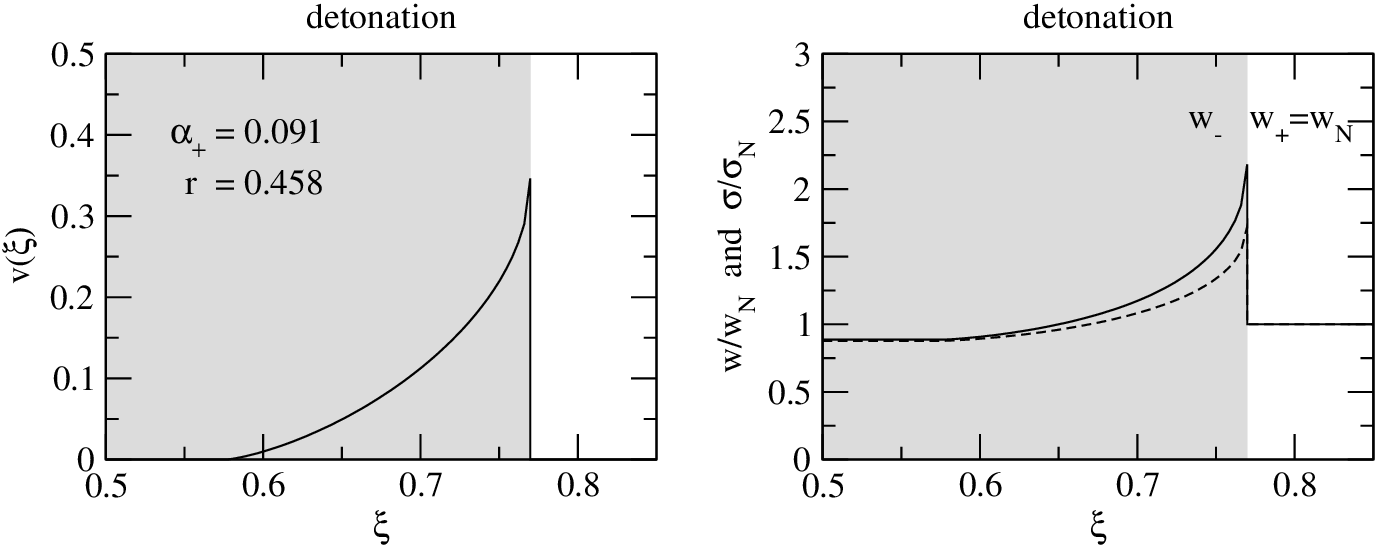}
\vskip 0.3 cm
\end{center}
\vskip -0.7 cm
\caption{\label{fig:expTFH}
\small Examples of the fluid velocity (in the plasma rest frame), enthalpy  
and entropy profiles for a subsonic
deflagration, a deflagration with rarefaction wave
(hybrid) and a detonation, for $a_-/a_+=0.85$. The bubble of broken phase 
is in gray. For  detonations, the fluid kinetic energy and thermal energy  
are concentrated  near the wall but behind it i.e. inside the bubble, 
while they are located  outside (mostly outside) of the bubble for 
deflagrations (hybrids). }
\end{figure}

A pictorial representation of a typical detonation is depicted
in Fig.~\ref{fig:bubbles}, right plot. The corresponding velocity profile 
is as in Fig.~\ref{fig:expTFH}, lower left plot. More precisely,
in detonations the phase transition wall moves at supersonic speed $\xi_w$
($\xi_w>c_s^+$) hitting fluid that is at rest in front of the wall. In the
wall frame, the symmetric-phase fluid is moving into the wall at
$v_+=\xi_w$ and entering the broken phase behind the wall where it slows
down so that $v_-<v_+$. In the rest frame of the bubble center, the fluid
velocity right after the wall passes jumps to $v(\xi_w) = \mu(v_+,v_-)$
(the Lorentz transformation (\ref{lorentz}) from the frame of the wall to
the rest frame of the center of the bubble)  and then slows down until it
comes to a stop, at some $\xi<\xi_w$, forming a rarefaction wave behind
the wall. From the previous discussion we know that $v$ will go to zero
smoothly at $\xi=c_s^-$. 

In order to obtain a consistent solution in the region $c_s^- < \xi <
\xi_w$, one needs $0 < \partial_{\xi} v < \infty$ which, using
eq.~(\ref{eq:flow}), requires $\mu (\xi) > \mu (\xi_w) \geq c_s^-$ behind
the wall. Consequently, detonation solutions are confined to the lower
right corner of fig.~\ref{fig:prof}, as indicated. Boosting to the wall
frame this implies $v_- \geq c_s^-$, since $v_- = \mu(\xi_w, v(\xi_w))$.
Therefore, detonations can be divided into Jouguet detonations
($v_-=c_s^-$) and weak detonations ($v_- > c_s^-$); strong detonations
($v_- < c_s^-$) are not consistent solutions of the fluid equations, see
fig.~\ref{fig:vpvm_contours}.\footnote{As $c_s^-$ can be different from
$1/\sqrt{3}$ in the most general case, the forbidden region $v_-<c_s^-$,
shaded in Fig.~\ref{fig:vpvm_contours}, will be shifted in those cases.}

Fig.~\ref{fig:expTFH} shows also the enthalpy profile (bottom right) for a 
detonation. Concerning this profile,
remember that the matching conditions across the wall give
\be
w_N = w_+ = w_- \left(\frac{1-\xi_w^2}{\xi_w}\right)
\left(\frac{v_-}{1-v_-^2}\right)\ , 
\ee
where the subscript $N$ denotes the plasma at the temperature of
nucleation far in front of the wall. Then, eq. (\ref{eq:enthalpy})
transforms into
\be
\label{wacross}
w(\xi) = w_N \left(\frac{\xi_w}{1-\xi_w^2}\right)
\left(\frac{1-v_-^2}{v_-}\right) \, 
\exp \left[ - 
\int_{v(\xi)}^{v(\xi_w)} \left(1+\frac{1}{c_s^2}\right)\gamma^2 \, \mu \, 
dv \right]\ .
\ee
Similar formulas can be derived for other quantities like the entropy 
(also shown in fig.~\ref{fig:expTFH}), the temperature, etc. It is 
straightforward to show that
\begin{equation}
T_- >T_+ = T_N \ .
\end{equation}
by using the detonation condition $r<1/(1+3\alpha_+)$.

\subsection{Deflagrations}

A pictorial representation of a typical deflagration is depicted in
Fig.~\ref{fig:bubbles}, left plot. The corresponding velocity profile is
as in Fig.~\ref{fig:expTFH}, upper left plot.  In contrast with
detonations, in deflagrations the plasma is at rest right behind the wall,
so that the wall velocity is now $\xi_w=v_-$. These solutions correspond
to the lower branches in Fig.~\ref{fig:vpvm_contours}, with a fluid
velocity that is larger behind the wall than in front, $v_->v_+$. From
Fig.~\ref{fig:vpvm_contours} we also see that in this case the
hydrodynamic relations across the wall imply $v_+ < c_s $. The fluid
velocity just in front of the wall jumps to $v(\xi_w) = \mu(v_-, v_+)$. As
$v_->v_+$, we get $v(\xi_w)<\xi_w$, so that the profile of deflagration
solutions start below the line $v=\xi$, as indicated in
fig.~\ref{fig:prof}.

As one moves out in $\xi$, $v(\xi)$ decreases and eventually would
become double-valued before reaching zero\footnote{In spite of
appearances, this happens even for very low values of the starting
$v(\xi)$ due to the improper-node nature of the point $(c_s,0)$ (see
discussion above).}.  Now we cannot accommodate the required jump to
zero velocity using the phase transition front, which is already fixed
at the beginning of the velocity profile. The way out of this dilemma
is that the flow can drop to zero in a shock-front: we can use
eq.~(\ref{eq:vvs}) where now quantities with subscript $\pm$ denote
the two sides of the shock-front instead of the phase transition
wall. (Notice also that the shock occurs in the symmetric phase where
in general we can assume the bag EoS holds.) At the shock-front there
is no discontinuity in $\epsilon$, and therefore $\alpha_+=0$, yet a
discontinuity in velocities is possible. As given by
eq.~(\ref{eq:vvs}) and shown in Fig.~\ref{fig:vpvm_contours}, this
requires~\cite{KurkiSuonio:1984ba} that the inward and outward
velocities of the fluid in the shock-front rest-frame fulfill the
relation $v_+ v_- = 1/3$, which in the plasma rest frame translates to
$\mu(\xi_{sh},v_{sh})
\, \xi_{sh}= c^2_s$. This condition determines the position of the shock 
front, which can be seen to occur before the singular point $\mu(\xi, 
v(\xi))=c_s$ (where $dv/d\xi\rightarrow -\infty$) is reached, see 
fig.~\ref{fig:prof}. 

Fig.~\ref{fig:expTFH} also shows the enthalpy and entropy profiles (top
right) for a deflagration. Notice that the temperature $T_+$ in this case
does not coincide with the temperature of the plasma outside the shock
$T_N$, since thermodynamic quantities are changing in front of the wall
and also at the shock front discontinuity. [For the same reason, explicit
formulas like (\ref{wacross})  would be now more involved.] As illustrated
by the plots in Fig.~\ref{fig:expTFH} one always has the inequality
$\omega_+>\omega_N$, i.e. $a_+T_+^4>a_N T_N^4$.  In fact, for
deflagrations one always has the inequalities 
\begin{equation} 
T_{+} > T_{sh}>T_N>T_- 
\end{equation} 
although the last one depends on the details of the Higgs equation of
motion, discussed in sect.~\ref{sec_static}. Hence the 
limit $\alpha_+<1/3$, that one found for $\alpha_+ = \epsilon /(a_+T_+^4)$ 
in deflagrations, translates into a weaker bound on the ratio $\alpha_N = 
\epsilon /(a_N T_N^4)$ as $\alpha_N>\alpha_+$.

An upper limit on $\alpha_N$ will be important later on to set an
upper limit on $\eta$. How large can $\alpha_N$ be for a fixed wall
velocity? From $\alpha_+<1/3$ we get the upper bound
$\alpha_N<\omega_+/(3\omega_N)$ and the ratio $\omega_+/\omega_N$
cannot be made arbitrarily large. Inspection of the deflagration
profiles in Fig.~\ref{fig:prof} shows that $\omega_+/\omega_N$ will be
maximized by the case with strongest shock-front, which corresponds to
the highest possible $v(\xi_w)$, that is, $v(\xi_w)=\xi_w$. In the
wall frame this gives $v_+=0$ and, using the matching conditions
(\ref{eq:wall_constr}), this case also has $\omega_-\rightarrow 0$,
{\it i.e.} $a_-T_-^4\rightarrow 0$. Physically this would represent a
limiting case for which the transition is such that the broken phase
inside the bubble is empty: the plasma is swept away by the wall (thus
leading to the strongest possible shock-front) and larger values of
$\eta$ cannot be realized microscopically. The same matching
conditions also give us $\omega_+
\rightarrow 4\epsilon$, which represents a fixed upper bound for
$\omega_+$. Going now to the shock rest frame and matching there one
gets $\omega_N=\omega_{sh}c_s^2(1-\xi_{sh}^2)/(\xi_{sh}^2-c_s^4)$.
The quantities $\omega_{sh}$ and $\xi_{sh}$ can be obtained once the
boundary condition $v(\xi_w)=\xi_w$ is fixed and lead to the minimal
value of $\omega_N$. The resulting upper bound on $\alpha_N$ can be
fitted numerically as a function of the wall velocity, and one gets
\begin{equation}
\left . \alpha_N^{max}\right |_{\mbox{\tiny defla.}}  \simeq 
\frac{1}{3}\frac{1}{(1-\xi_w)^{-13/10}}\ ,
\end{equation}
as derived in Appendix~\ref{sec_fits}.
\begin{figure}[htb!]
\begin{center}
\includegraphics[width=0.65\textwidth, clip ]{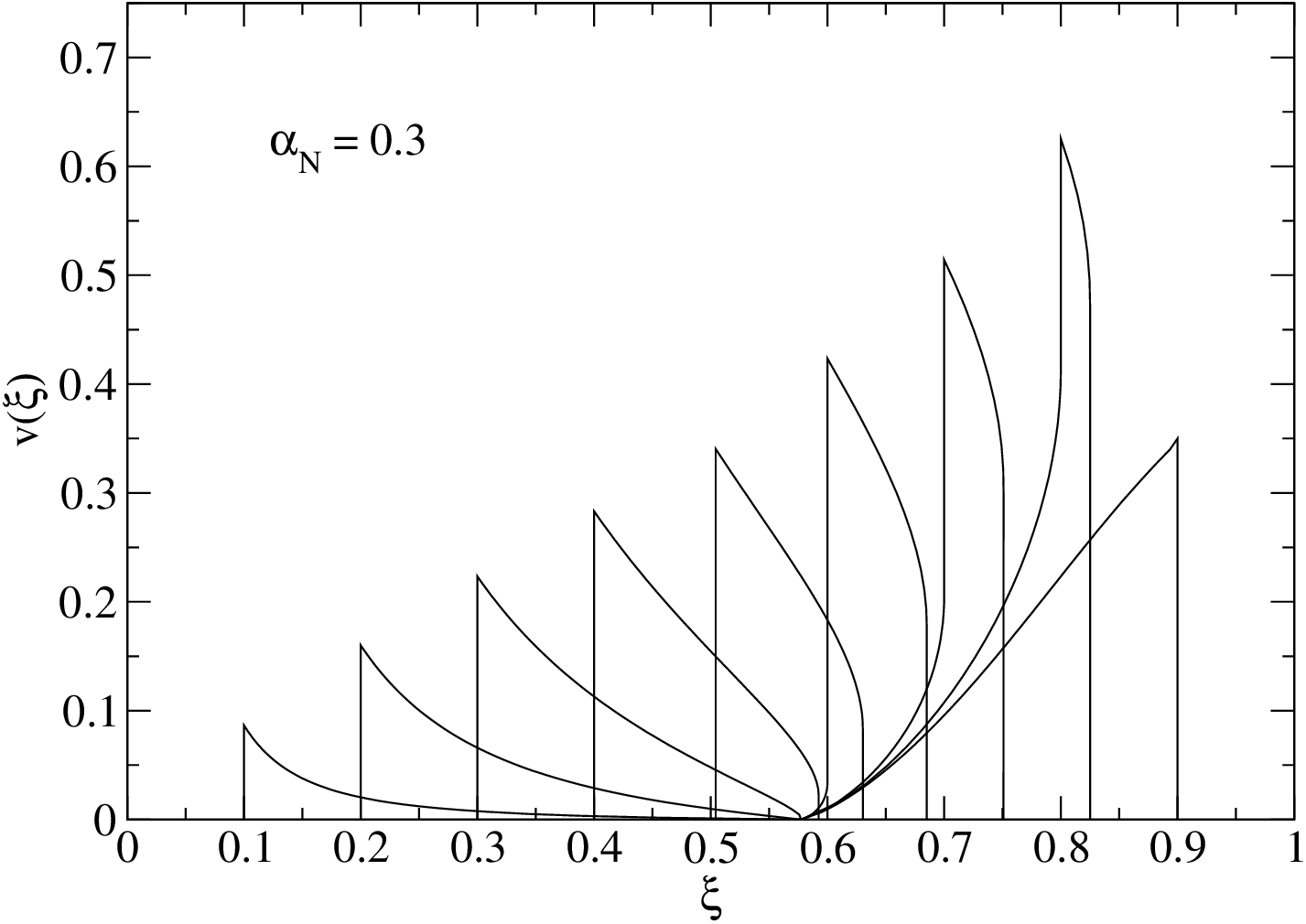}
\caption{\label{fig:expAllinOne}
\small Fluid velocity profiles  (in the plasma rest frame) 
for deflagrations, hybrids and detonations, for different wall velocities 
and $\alpha_N=0.3$. }
\end{center}
\end{figure}

\subsection{Hybrids}
\label{subsec:hybrids}

From the previous discussion of velocity jumps it is clear that it should
be possible to combine detonation and deflagration solutions into a new
velocity profile that is a superposition of both types (thus the name
hybrid), provided the wall is supersonic. In fact, it is known from
hydrodynamics simulations~\cite{KurkiSuonio:1995pp} that supersonic
deflagrations are not stable but develop a rarefaction wave, identical to
the detonation profile discussed earlier. A pictorial representation of a
typical hybrid bubble is depicted in Fig.~\ref{fig:bubbles}, central plot.
The corresponding velocity profile is shown in Fig.~\ref{fig:expTFH}, middle
left plot.  Applying the hydrodynamic constrains for deflagration and
detonation solutions we have (see fig.~\ref{fig:vpvm_contours}) $\mu
\left(\xi_w,\, v(\xi_w^-)\right) = v_- \geq c_s^-$ and $\mu \left(\xi_w,\,
v(\xi_w^+)\right) = v_+ \leq c_s^+$.  On the other hand, entropy
considerations enforce $v_- \leq c_s^-$, so that the rarefaction wave has
to be of Jouguet type, with $v_-=c_s^-$.

Therefore in these hybrid solutions the wall velocity $\xi_w$ is not
identified with either $v_+ $ or $v_-$, and the phase transition front
is followed by a rarefaction wave of Jouguet type
($\xi_w>v_-=c_s^->v_+$) and is preceded by a shock front. Typical
enthalpy and entropy profiles are shown in the middle right plot of
Fig.~\ref{fig:expTFH}. As the wall velocity increases, the
deflagration part of the solution becomes thinner, and eventually
disappears, as shown in Fig.~\ref{fig:expAllinOne}.
\vspace{0.5cm}
\begin{figure}[htb!]
\begin{center}
\includegraphics[width=0.65\textwidth, clip ]{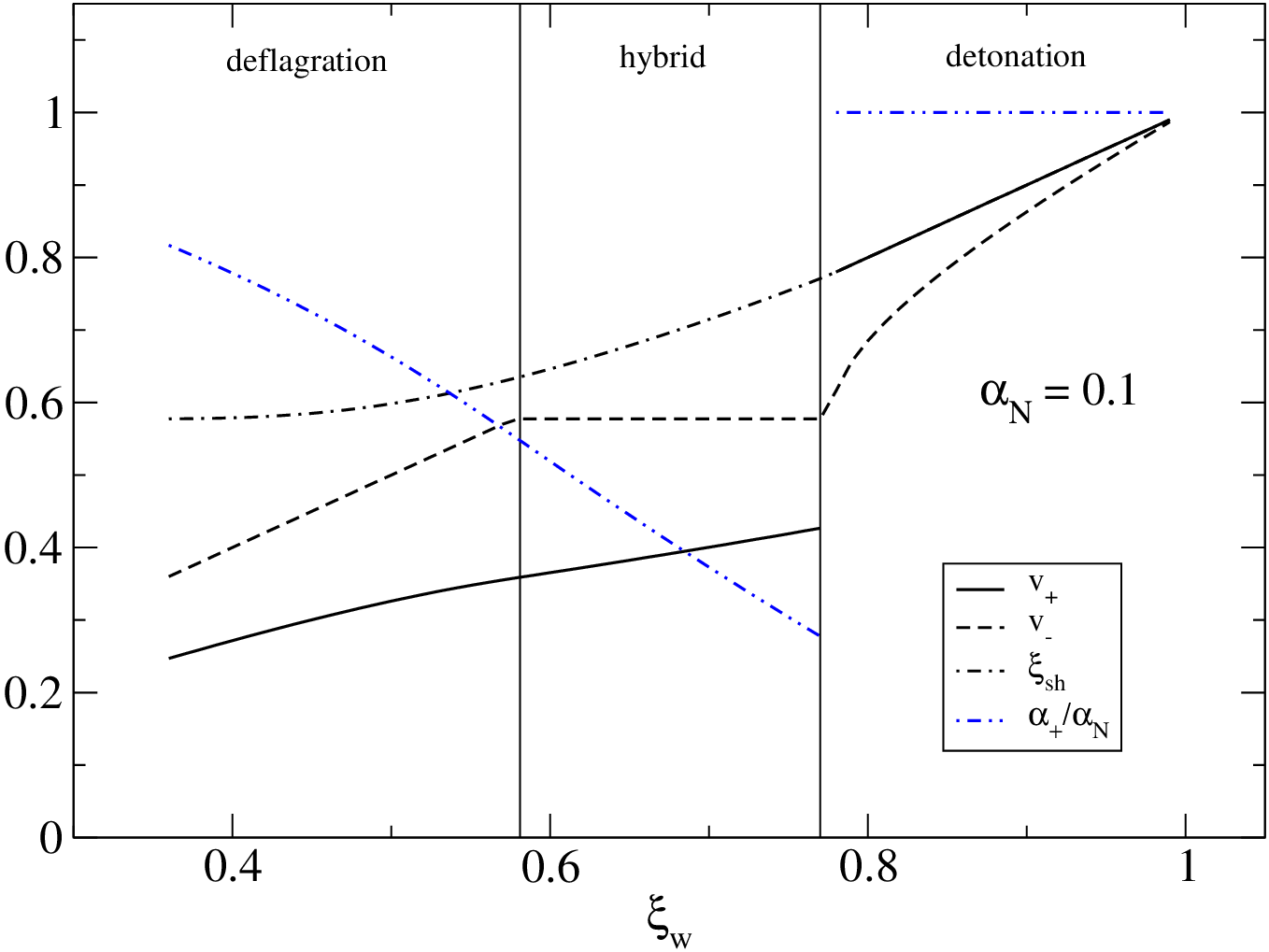}
\caption{\label{fig:expAll}
\small The flow velocities in the wall frame, $v_+$ and $v_-$, the velocity
of the shock front $\xi_{sh}$ and the ratio $\alpha_+/\alpha_N$ as a
function of the wall velocity (for $\alpha_N=0.1$).}
\end{center}
\end{figure}
\begin{figure}[ht!]
\begin{center}
\includegraphics[width=0.8\textwidth, clip ]{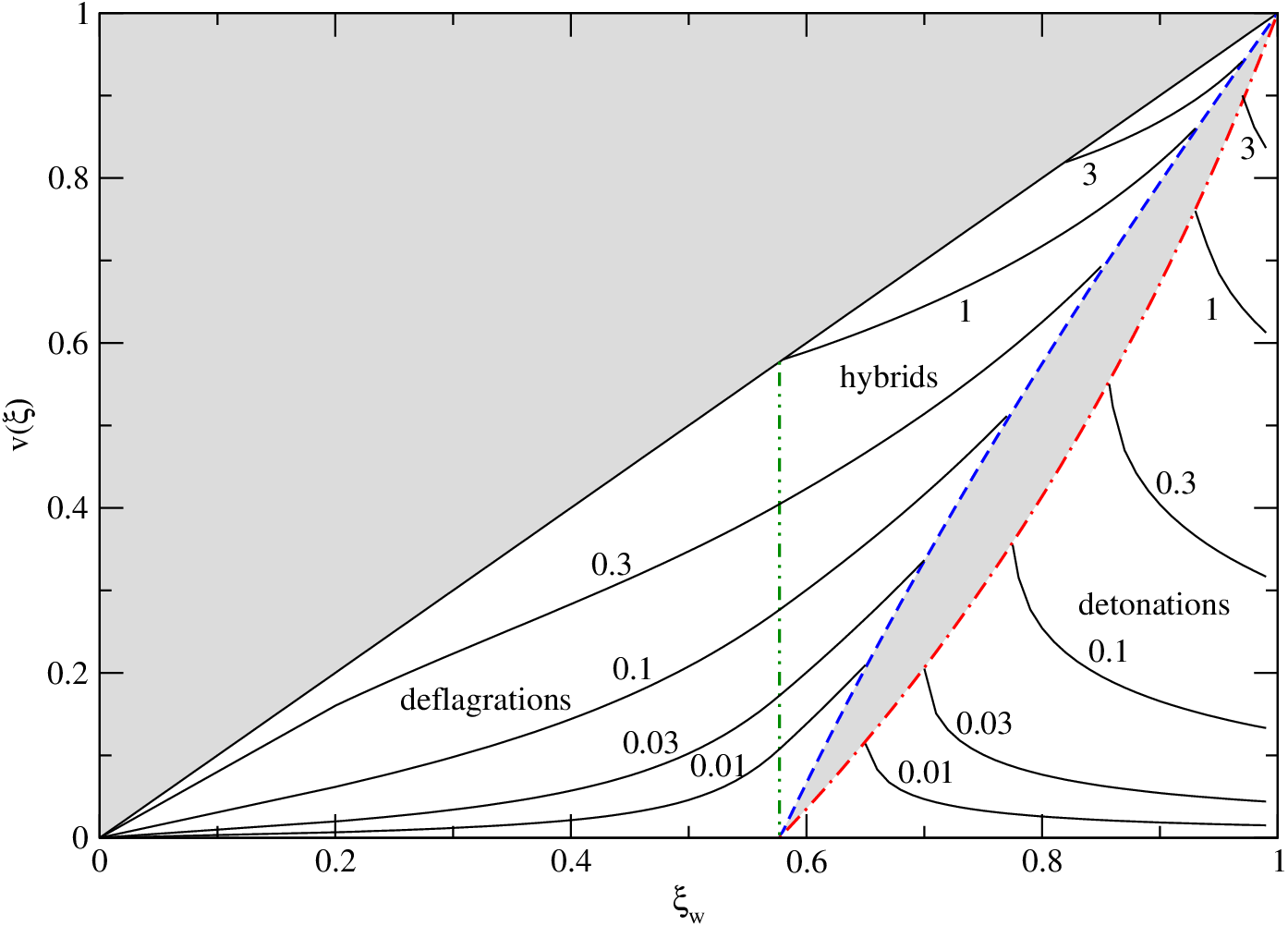}
\caption{\label{fig:flow_max}
\small The maximal flow velocity in the plasma rest frame as a function of the
wall velocity and several values of $\alpha_N$, as indicated by the 
labels of each curve. }
\end{center}
\end{figure}

In summary, one obtains the following picture. For subsonic wall
velocities, the bubble expansion proceeds in general by deflagrations.
If for constant $\alpha_N$ the wall velocity is increased beyond the
velocity of sound, the flow profile develops a rarefaction wave. If
the wall velocity is further increased, the shock becomes thinner
until it completely vanishes and the bubble expansion proceeds by a
Jouguet detonation. When the shock front vanishes, some of the
quantities right in front of the wall experience a jump, namely the
temperature $T_+$, the flow velocity $v_+$ and also $\alpha_+$ as
illustrated in Fig.~\ref{fig:expAll}. Beyond this point the solutions
are given by weak detonations ($v_->c_s^-$). The hybrid solutions then
fill the gap of wall velocities between Jouguet deflagrations and
Jouguet detonations~\cite{KurkiSuonio:1995pp}, as seen in
Fig.~\ref{fig:flow_max} which gives the maximal flow velocity in the
plasma rest frame as a function of the wall velocity $\xi_w$ for fixed
values of $\alpha_N$. For deflagrations and hybrids the maximal flow
is in front of the wall while for detonations it is behind it. Notice
that, in the transition from hybrids to detonations, the maximal flow
velocity jumps without a gap in $\xi_w$. The whole flow profile can be
reconstructed using Fig.~\ref{fig:prof}.

\section{Efficiency coefficients\label{sec_eff}}

In the following we present the analysis of the efficiency
coefficient $\kappa$ (the ratio of bulk kinetic energy to vacuum 
energy) following ref.~\cite{Landau, KurkiSuonio:1984ba,
Kamionkowski:1993fg}. Steinhardt argued in his seminal 
work~\cite{Steinhardt:1981ct} that
the relative plasma velocity behind the wall should just be given by
the speed of sound, $v_-=c_s^-$. Using this piece of information
in eq.~(\ref{eq:vvs}) gives immediately
\be
\xi_w = \xi_J \equiv \frac{\sqrt{ \alpha_+ (2 + 3 \alpha_+ )} + 1}
{ \sqrt{3} (1 + \alpha_+)} \quad \textrm{(Jouguet detonations)},
\ee
and
\be
\label{eq:JDetKappa}
\kappa \simeq \frac{\sqrt{\alpha_N}}{0.135 + \sqrt{0.98 + \alpha_N}}
\quad \textrm{(Jouguet detonations)}.
\ee
Notice that this differs from the result in ref.~\cite{Kamionkowski:1993fg}
due to our different definition of the efficiency factor $\kappa$ [see 
discussion after eq.~(\ref{eq:defKap})]. 

\begin{figure}[ht]
\begin{center}
\includegraphics[width=0.65\textwidth, clip ]{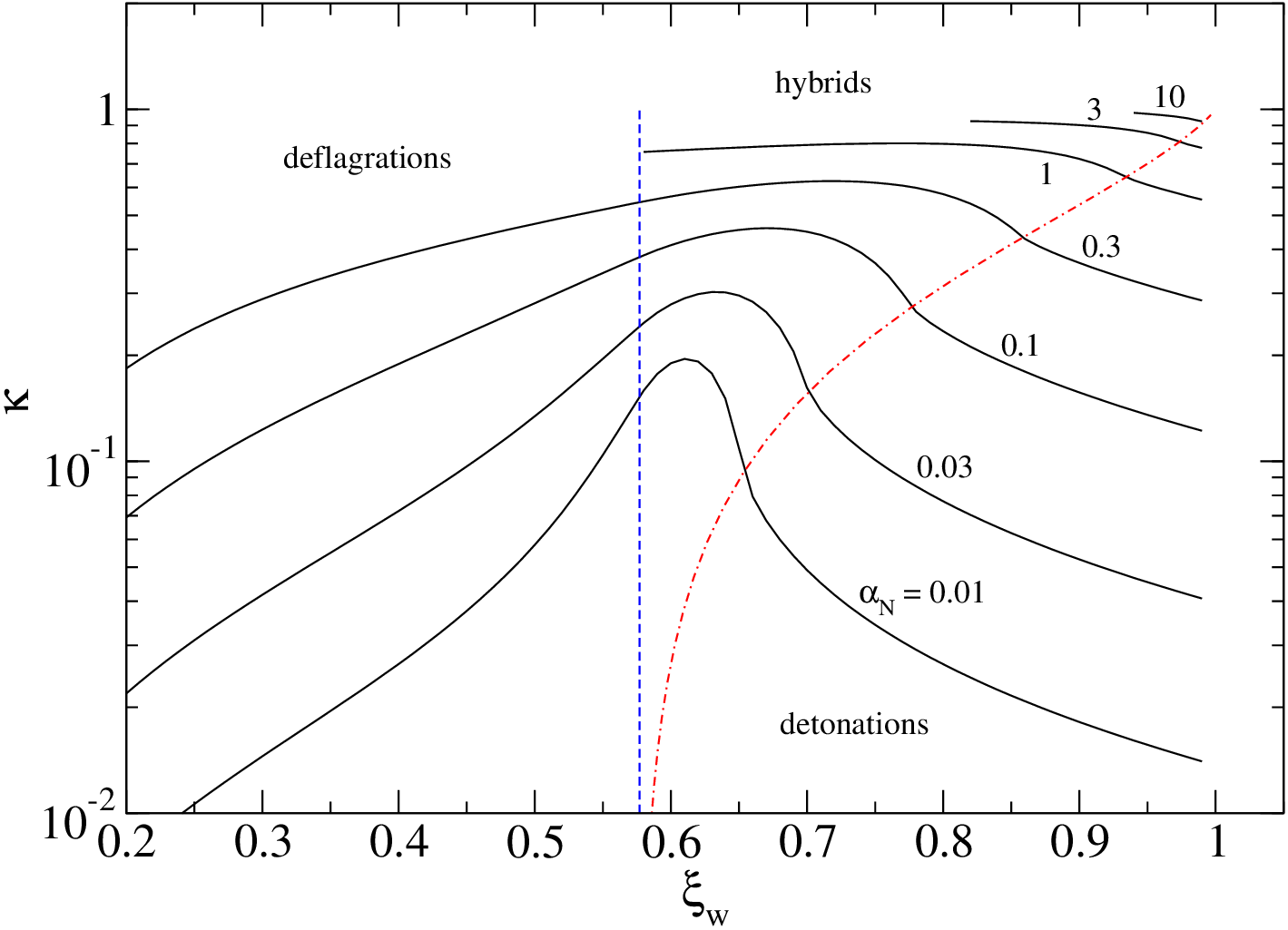}
\caption{\label{fig:KfuncH}
\small The efficiency coefficient $\kappa$ as a function of the wall velocity
$\xi_w$ for fixed $\alpha_N$. The dashed and dashed-dotted lines mark 
the transitions from deflagrations to hybrids and further to 
detonations. The dashed-dotted line corresponds to Jouguet detonations 
(the only case used in the literature, although with a missing 
$1/\xi_w^3$ factor). Analytical fits for $\kappa(\alpha_N,\xi_w)$ are 
provided in Appendix ~\ref{sec_fits}.}
\end{center}
\end{figure}

As already mentioned, the Jouguet condition turns out to be unrealistic in
cosmological phase transitions~\cite{KurkiSuonio:1995pp}, and therefore
one needs to rederive a formula for $\kappa$ that supersedes
eq.~(\ref{eq:JDetKappa}). In the next section we will discuss how to
relate the two velocities in the plasma $v_+$ and $v_-$ (which is
equivalent to the determination of the bubble wall). Knowing the wall
velocity $\xi_w$ and the parameter $\alpha_+$ (or $\alpha_N$) the velocity
profile is determined and $\kappa$ can be calculated using
eq.~(\ref{eq:defKap}) independently from further assumptions on friction
and microscopic physics in the plasma close to the wall (which are
relevant to fix $\xi_w$).

The results are shown in Fig.~\ref{fig:KfuncH} which gives $\kappa$ as
a function of the wall velocity for several values of the vacuum
energy $\alpha_N$. Note also that for large values of $\alpha_N$,
small wall velocities are impossible, see Fig.~\ref{fig:flow_max} and
the discussion about deflagrations in Sec.~\ref{sec_detdef}. The
efficiency increases with $\alpha_N$ and is maximal for the hybrid
solutions. Nevertheless, according to numerical simulations, the
detonation solutions are the only supersonic modes that are globally
stable for small values of $\alpha_N$ and realistically the maximal
efficiency corresponds to the Jouguet case in this regime. The gravity
wave literature focused on the Jouguet detonations (dashed-dotted
line) and hence overestimated the efficiency $\kappa$. However,
we stress that this effect is mostly compensated by the missing factor
$\xi_w^3$ in the formula of $\kappa(\alpha)$ we mentioned before.  In
appendix~\ref{sec_fits}, we give fits to the efficiency $\kappa$ shown
in Fig.~\ref{fig:KfuncH} as a function of the parameters $\alpha_N$
and $\xi_w$.
\begin{figure}[ht]
\begin{center}
\includegraphics[width=0.65\textwidth, clip ]{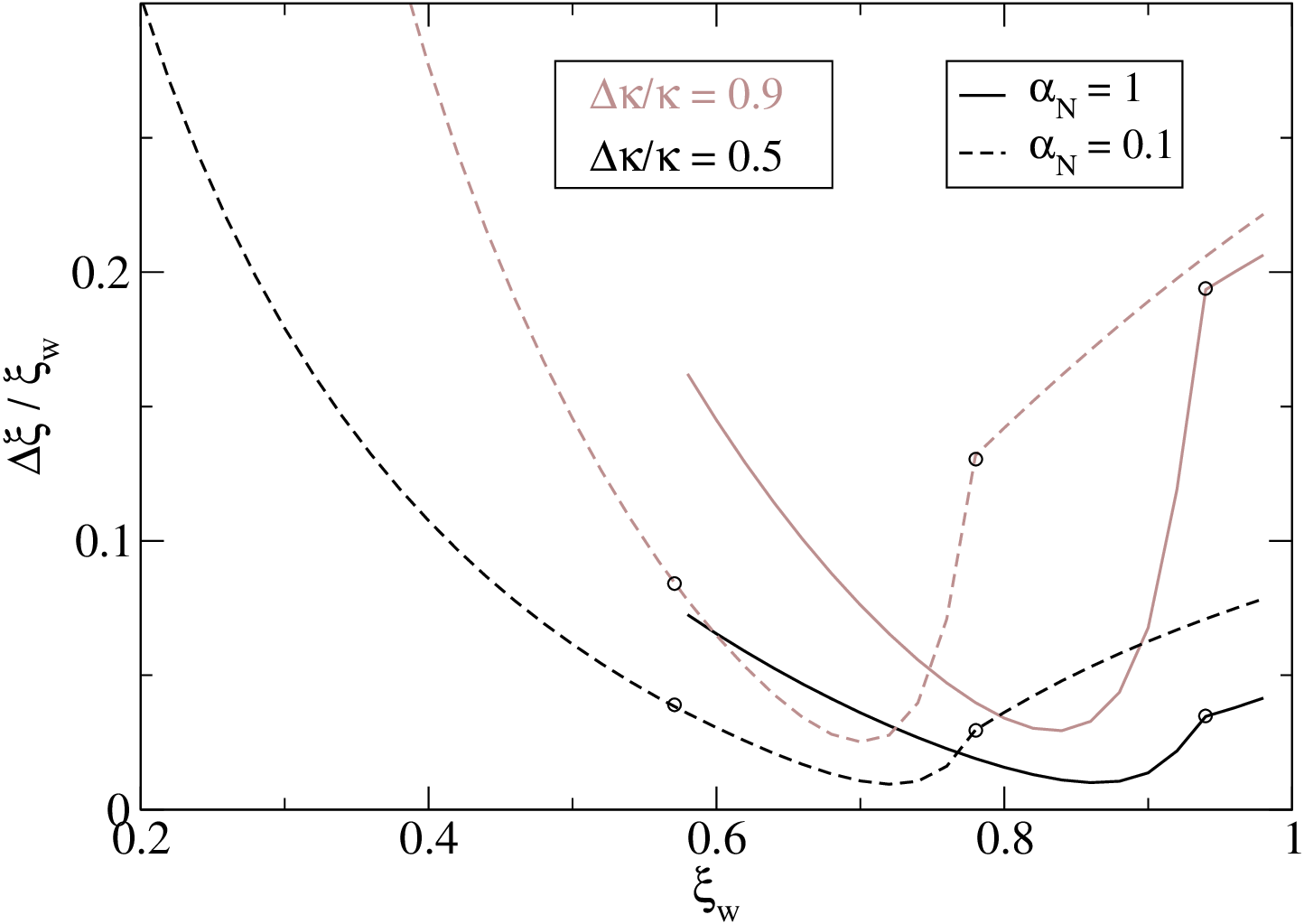}
\caption{\label{fig:thick}
\small Curves with a fixed fraction of kinetic energy 
$\Delta\kappa/\kappa$ in a shell of thickness $\Delta \xi$ for 
$\alpha_N=0.1$ and
$\alpha_N=1$ as a function of the wall velocity $\xi_w$. The lower black
(upper brown) lines denote $\Delta\kappa/\kappa=0.5$ 
($\Delta\kappa/\kappa=0.9$). The dots in each
line mark the boundaries between different regimes (deflagrations, hybrids
and detonations, in order of increasing wall velocity).}
\end{center}
\end{figure}

Finally, it is interesting to estimate the thickness of the plasma shell
near the bubble wall where the kinetic energy in the plasma is
concentrated (as this is relevant for GW production). In
Fig.~\ref{fig:thick} we give the thickness $\Delta \xi$ of a shell around
the transition wall such that it contains a given fraction of the kinetic
energy, $\Delta \kappa/\kappa$, as indicated. For each type of bubble we
choose the shell so as to maximize that fraction. That is, for detonations
the shell is $(\xi_w,\xi_w-\Delta\xi_w)$; for deflagrations
$(\xi_w-\Delta\xi_w,\xi_w)$ and in-between for hybrids. The dots in each
line mark the boundaries between different regimes (deflagrations, hybrids
and detonations, in order of increasing wall velocity). We see that,
especially for hybrid solutions, the kinetic energy is localized in the
shock, which can be rather thin in this case. For weaker deflagrations,
the shock is thicker. For detonations, part of the kinetic energy is
always in the tail of the rarefaction wave.

In the limit of ultra-relativistic wall velocities ($\xi_w\rightarrow 1$)
the velocities $v_\pm$ approach 1 with a relative slope $v_+/v_-$ that
depends on $\alpha_+$ (see Fig.~\ref{fig:vpvm_contours}). Generically, the
fluid velocity will not be ultra-relativistic in the plasma frame and the
bubbles will expand as detonations (see Fig.~\ref{fig:prof}). Expanding
(\ref{eq:vvs2}) in powers of $1/\gamma_\pm$ one gets
\be
\label{eq:highv1}
\frac{\gamma_+^2}{\gamma_-^2} = 1 + 3 \alpha_N\ , \quad
v(\xi_w) = \mu(v_+, v_-) = \frac{3 \alpha_N}{ 2 + 3 \alpha_N}\ ,
\ee
where we have used $\alpha_+=\alpha_N$ as we are in the regime of 
detonations. This relativistic case requires $r=1/(1+3\alpha_+)$, which 
translates into $p_+-p_-=2\epsilon$.
Numerically, the resulting efficiency in this limit is given by
\be
\label{eq:kapBMstat}
\kappa \simeq \frac{\alpha_N} {0.73 + 0.083 \sqrt{\alpha_N} +  \alpha_N}
\quad \textrm{(relativistic $\xi_w$)}.
\ee
This result is relevant for the analysis of the case of runaway walls
in section \ref{sec_runaway}. 

\section{Bubble wall velocity for steady state walls\label{sec_static}}

When a bubble of the broken phase is nucleated, and is large enough to
start growing, it will expand in an accelerated way, with the difference
in free-energy across its wall acting as driving force. There is however a
resistance to this expansion from the surrounding plasma, which exerts a
friction force that grows with the velocity of the moving wall. 
Eventually,
an equilibrium between these two forces is reached after a short time of
expansion and, since then on, the bubble wall keeps expanding in a steady
state at a constant terminal velocity. As explained in the last sections,
hydrodynamics alone cannot be used to determine this terminal wall
velocity and one has to analyze the mechanism of entropy production and
friction in the wall.

\subsection{EoM for the Higgs field and the 
friction parameter $\eta$}

We take into account entropy production and friction through the equation 
of motion of the Higgs 
field 
\be
\label{HEOMprior}
\square \phi + \frac{\partial {V_0}}{\partial \phi} 
+\sum_i \frac{dm_i^2}{d\phi} \int \frac{d^3p}{(2\pi)^32E_i} f_i(p)= 0 \ .
\ee
 By decomposing 
\be
 f_i(p)=f_i^{eq}(p)+\delta f_i(p)\ ,
\ee
where $f_i^{eq}=1/[\exp{(E_i/T)}\mp 1]$ is the equilibrium distribution 
function of
particle species $i$ with $E_i^2=p^2+m_i^2$, eq.~(\ref{HEOMprior}) takes 
the simple form (see also ref.~\cite{Ignatius:1993qn} and more recently
ref.~\cite{Megevand:2009ut})  
\be
\label{HEOM}
\square \phi + \frac{\partial {\cal F}}{\partial \phi} 
- {\cal K}(\phi) = 0 \ ,
\ee
where the second term gives the force driving the wall and
 ${\cal K}(\phi)$ stands for the friction term 
 \be
 {\cal K}(\phi)=-\sum_i \frac{dm_i^2}{d\phi} \int 
\frac{d^3p}{(2\pi)^32E_i} \delta f_i(p)\ .
 \ee

Friction is therefore due to deviations of particle distributions from
equilibrium. In principle, calculation of ${\cal K}(\phi)$ requires
solving a coupled system involving Boltzmann equations for particle
species with a large coupling to the Higgs field. This intricate
calculation has been performed in the Standard Model
\cite{Moore:1995si} and in the MSSM \cite{John:2000zq} and under the
assumption that the deviation from thermal equilibrium is small,
i.e. $ \delta f_i(p)\ll f_i(p)$, which is only true for weakly
first-order phase transitions.
 
In this paper, we want to follow a more phenomenological 
and model-independent approach.
In refs.~\cite{Ignatius:1993qn, Megevand:2009ut} a particularly simple
choice for ${\cal K}(\phi)$ was used:
\be
\label{eq:Ksimple}
{\cal K}(\phi) = T_N \, \tilde \eta \, u^\mu \partial_\mu \phi\ ,
\ee
(where $T_N$ is inserted just to make $\tilde\eta$ dimensionless).
This Lorentz invariant choice is motivated by similar approaches in the
inflationary context but, as we will see in the next section, it does
not lead to the correct behavior for highly relativistic bubble wall
velocities: this friction force could
increase without bounds, due to the $\gamma$ factor appearing through
$u^\mu \partial_\mu\phi$, but we know from
ref.~\cite{BM} that at large wall velocities the friction
term approaches a constant (see next section).  

Friction comes from out-of-equilibrium effects and the assumption that
it depends locally only on the plasma four-vector $u^\mu$ and a
Lorentz scalar $\eta$ is too simplistic.  In our phenomenological
approach we ensure that the friction force grows with $v$ and not
$\gamma v$. Such behavior could arise from a friction term in the
Higgs equation of motion of the form
\be
\label{eq:Ksimple2}
{\cal K}(\phi) = T_N \, \tilde \eta \, \frac{ u^\mu \partial_\mu \phi}
{\sqrt{1+ (\lambda_\mu u^\mu)^2}}\ ,
\ee
where the Higgs background is parametrized by a four-vector $\lambda^\mu$
[such that $\phi(\lambda^\mu x_\mu)$ and $\lambda^\mu$ is $(0,0,0,1)$
in the wall frame]. One can show that the entropy production from such 
a term is always positive, as it should be.

Assuming then that in the steady state the bubble is large enough 
so that we can use the planar limit, using (\ref{eq:Ksimple2}) in 
eq.~(\ref{HEOM}) we get, in the wall frame,
\be
\label{HEOMw}
\partial_z^2\phi-\frac{\partial{\cal F}}{\partial\phi}=-T_N\tilde{\eta} 
v \partial_z\phi\ ,
\ee
where $z$ is the direction of the wall velocity. Note that the right-hand 
side would be multiplied by $\gamma$ if we use (\ref{eq:Ksimple}) instead 
of (\ref{eq:Ksimple2}). If we multiply  this differential equation by 
$\partial_z\phi$ on both sides and integrate across the wall, we get
\be
\label{eq:MS_EoM}
\int \, dz \, \partial_z \phi\ \frac{\partial {\cal F}}{\partial \phi}
= T_N\tilde{\eta} \int \, dz \, v\ (\partial_z \phi )^2 \ .
\ee
The integration of the force term could be simply performed if the
free energy ${\cal F}$ did not have an implicit dependence on $z$
via the change in the temperature, $T(z)$, with $T(\pm\infty)=T_\pm$.
Using
$d{\cal F}/dz=(\partial {\cal F}/\partial \phi)\partial_z\phi+
(\partial {\cal F}/\partial T)\partial_z T$, one can rewrite 
the driving force of the bubble expansion as:
\be
\label{eq:force0}
F_{dr} \equiv \int \, dz \, \partial_z \phi\ \frac{\partial {\cal 
F}}{\partial \phi}
= \left.{\cal F}\right|_-^+
-\int \, dz \, \partial_z T\ \frac{\partial {\cal F}}{\partial T}
\ ,
\ee
and, using $\epsilon_\pm$ and $a(z)$ as defined in eqs.~(\ref{aepsgen})
and (\ref{eosg}), one gets, without making assumptions on the plasma
equation of state:
\be
\label{eq:force}
F_{dr} = \epsilon_+ - \epsilon_- - \frac13 \int \, da \,T^4\ .
\ee
By making further use of the definition of $a(z)$ and assuming that the
distribution functions for particle species are the equilibrium ones one
can rewrite eq.~(\ref{eq:force}) as
\be
\label{Fdriving}
F_{dr}=\Delta V_0+\sum_i |N_i|\int d z \frac{d m_i^2}{d z}\int 
\frac{d^3p}{(2\pi)^3}\frac{f_i^{eq}}{2E_i}\ ,
\ee
where $\Delta V_0$ is the $T=0$ 
part
of $\epsilon_+-\epsilon_-$, that is, the difference in $(T=0)$ potential 
energy between the symmetric and broken minima ($\epsilon$, for the bag 
equation of state). This expression for the
driving force  will be useful in sect.~\ref{sec_runaway}.

Notice  that this force does not coincide with the latent heat
$\Lambda$, given by
\be
\Lambda \equiv e_+ - e_- =
\left.\left(\epsilon+a\ T^4\right) \right|^+_-\ .
\ee
nor with the free energy (pressure) difference 
\begin{equation}
\Delta {\cal F} \equiv p_- - p_+ =
\left.\left(\epsilon-\frac{a}{3}\ T^4\right) \right|^+_-\ .
\label{eq:freeenergydiff}
\end{equation}
Using the fact that the number of effective degrees of freedom decreases
continuously in the wall, the last integral in (\ref{eq:force}) is bounded as
\be
\label{integr}
\frac13 (a_+ - a_-) T_{min}^4 \ \leq \ \frac13 \int \, da \,T^4 \ \leq\  
\frac13 (a_+ - a_-) T_{max}^4\ ,
\ee
where $T_{min}={\mathrm{Min}}\{T_-,T_+\}$ and $T_{max}={\mathrm{
Max}}\{T_-,T_+\}$.  For weak phase transitions one typically has $T_+
\approx T_-$ and the concrete choice of the profiles in the wall are not
important. For very strong
phase transitions, the free energy is dominated by the vacuum contribution
and the plasma contribution is rather small altogether:
\begin{eqnarray}
F_{dr} \approx   \Delta {\cal F} & \approx & \epsilon_+ -\epsilon_-  - 
(a_+-a_-) \frac{T^4}{3}  \ \ \ \mbox{(for weak phase transitions)}\\
F_{dr} \approx   \Delta {\cal F} \approx \Lambda  & \approx &  \epsilon_+ 
-\epsilon_-    \ \ \ \mbox{(for strong phase transitions)}
\end{eqnarray}

For simplicity, let us approximate the integral in eq.~(\ref{integr})
by the expression involving $T_+$. For other approximations the results would
not change qualitatively. In this case, the Higgs equation of motion
gives
\be
\label{eq:micro1}
\alpha_+ - \frac13\left(1 - \frac{a_-}{a_+}\right) = 
\frac{\tilde{\eta}T_N}{a_+T^4_+}\int dz\ v\ (\partial_z\phi)^2\ .
\ee
In the SM, the second term in (\ref{eq:micro1}) is roughly $(1- a_-/a_+)/3
= 0.05$. For small values of $\alpha_+$ the left hand side of 
(\ref{eq:micro1}) is negative and no
bubbles can nucleate (for nucleation one can consider $T_+=T_-$, in which
case $F_{dr}={\cal F}_+-{\cal F}_-$, so that the wrong sign of the 
free-energy
difference is the reason that prevents the phase transition). For larger
values of $\alpha_+$, the left-hand side of the equation is positive and
has to be balanced by the friction force so as to obtain a constant wall
velocity, as already explained. We will rewrite the right hand side of
(\ref{eq:micro1}) as
\be
\frac{\tilde{\eta}T_N}{a_+T^4_+}
\int dz\ v\ (\partial_z\phi)^2
\equiv \eta\frac{\alpha_+}{\alpha_N}
\langle v\rangle\ ,
\ee
which serves as the definition of $\eta$. Here $\left< v\right>$
denotes the fluid velocity average across the wall (in the wall frame), 
that we approximate
as
\be
\label{eq:def_mean}
\left< v \right> \equiv \frac{\int dz\ v\ (\partial_z\phi)^2}{\int dz\ 
(\partial_z\phi)^2}
\simeq  \frac12 (v_+ + v_-)\ .
\ee
In this way, (\ref{eq:micro1}) is simply written as\footnote{Compared
to ref.~\cite{Megevand:2009ut}, our  velocity
average does not include a $\gamma$ factor and we use a different expression for the driving force.}
\be
\label{eq:micro2}
\alpha_+ - \frac13\left(1 - \frac{a_-}{a_+}\right) = 
\eta \frac{\alpha_+}{\alpha_N}  \left< v \right>\ .
\ee

Eq.~(\ref{eq:micro2}) reproduces the correct behavior for small and very
large wall velocities, even though the parameter $\eta$ will not be the
same in these two limits.  This simple phenomenological approach already
reproduces almost all qualitative features found in the hydrodynamic
simulations performed in ref.~\cite{Ignatius:1993qn}. Notice that the
right hand side of (\ref{eq:micro1})  is an increasing function of the
wall velocity (since $v_+$ and $v_-$ are, see also Fig.~\ref{fig:expAll}).
On the other hand, the left hand side of the equation is decreasing for
small velocities and jumps when the flow solutions change from
deflagrations to detonations (also the velocity $v_+$ jumps but this is
less essential). Hence, for a certain range of values of the friction
coefficient $\eta$, eq.~(\ref{eq:micro1}) has two solutions while in
ref.~\cite{Ignatius:1993qn} three solutions were found. The discrepancy is
due to the fact that we neglect the thickness of the wall compared to the
shock and this would smooth out the jumps in $v_+$ and
$\alpha_+/\alpha_N$, thus producing a third solution. The three solutions
would in the present model then be a deflagration (or a hybrid solution),
a weak detonation and a configuration close to a Jouguet detonation, in
agreement with the results found in ref.~\cite{Ignatius:1993qn}.

Finally, at the critical temperature at which the two phases have
degenerate free energy, one expects that the gravitational wave signal
vanishes.  However, an analysis using the Jouguet condition still
implies supersonic bubble expansion. This led to the fact that the
difference in free energy is often confused with the vacuum energy
$\epsilon$ in the literature on gravitational wave production that is
based on the Jouguet condition. Replacing the Jouguet condition by the
equation of motion (EoM) of the Higgs field solves this problem
consistently, since at the critical temperature one obtains $\alpha_N
= (1 - a_-/a_+)/3$ and, according to equation (\ref{eq:micro2}), the
bubble cannot expand.

\subsection{Wall velocity in the  $(\eta,\alpha_N)$ plane}

\begin{figure}[htb!]
\begin{center}
\includegraphics[width=0.45\textwidth, clip ]{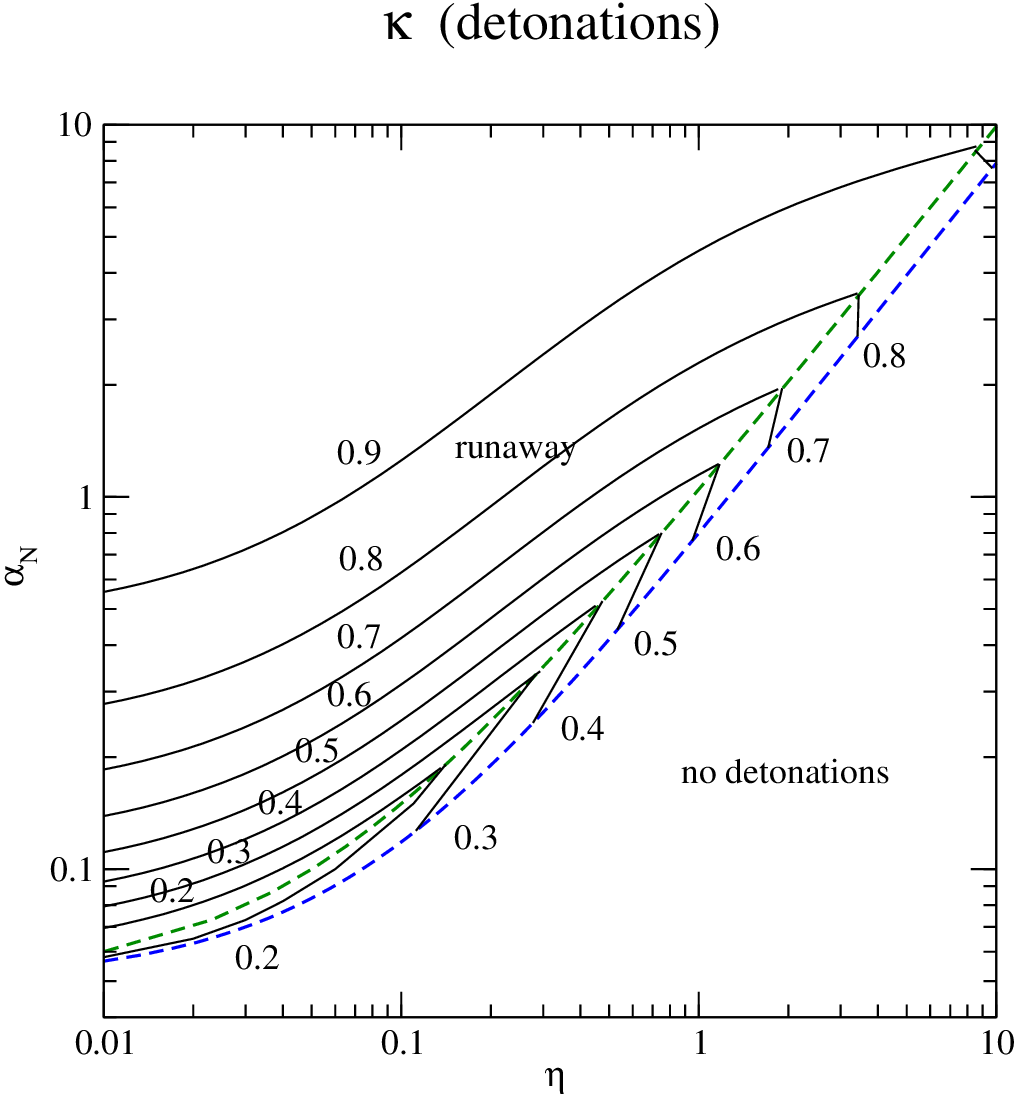}
\includegraphics[width=0.45\textwidth, clip ]{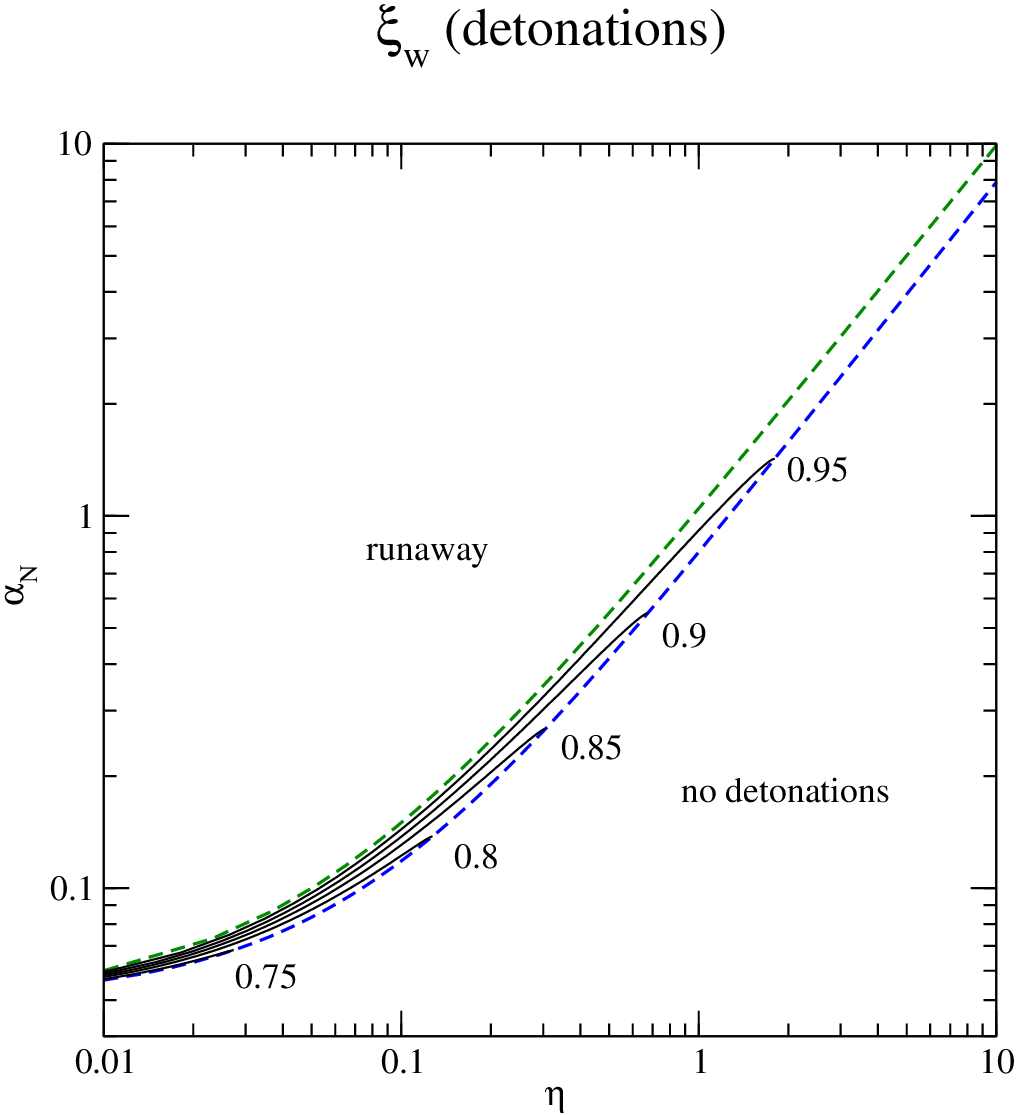}
\vskip 0.5 cm
\includegraphics[width=0.45\textwidth, clip ]{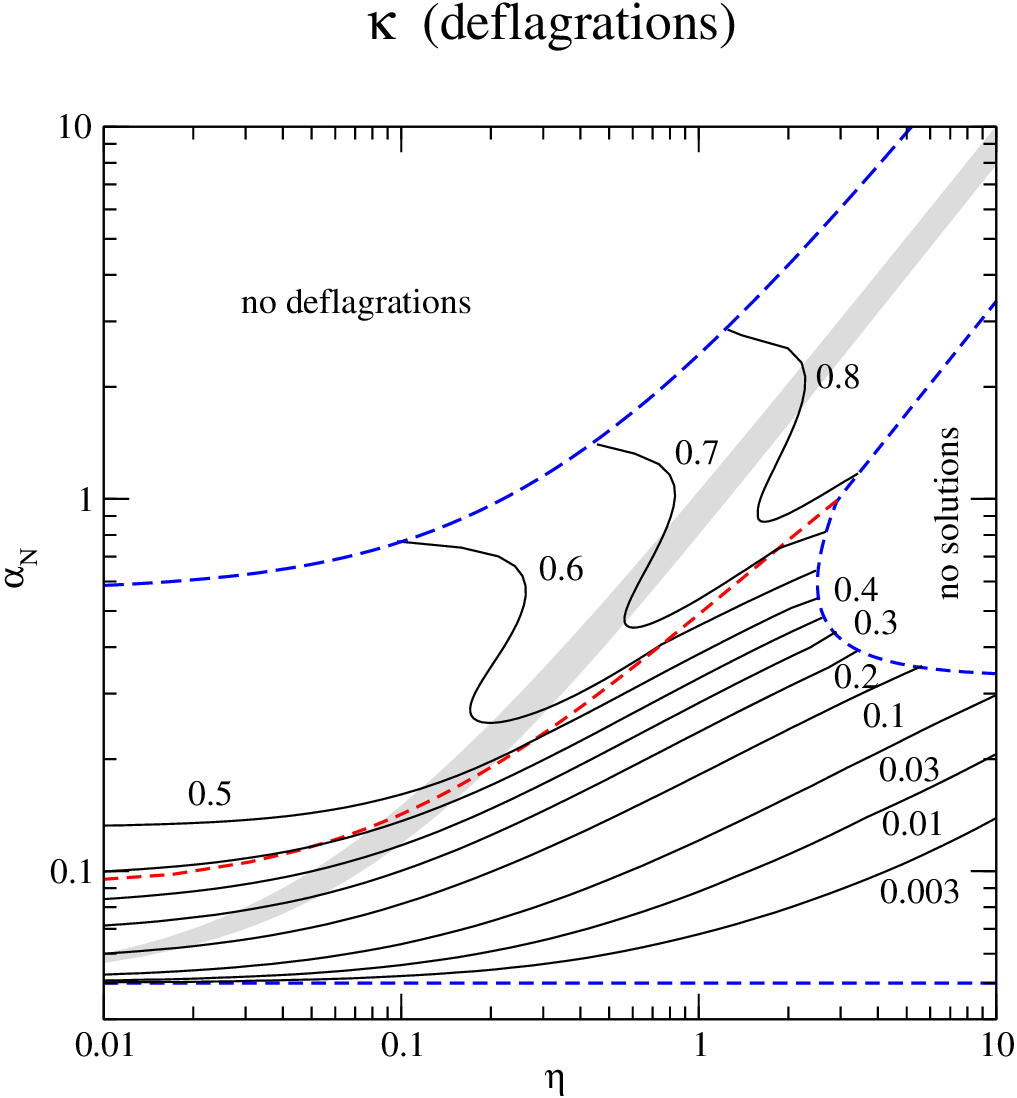}
\includegraphics[width=0.45\textwidth, clip ]{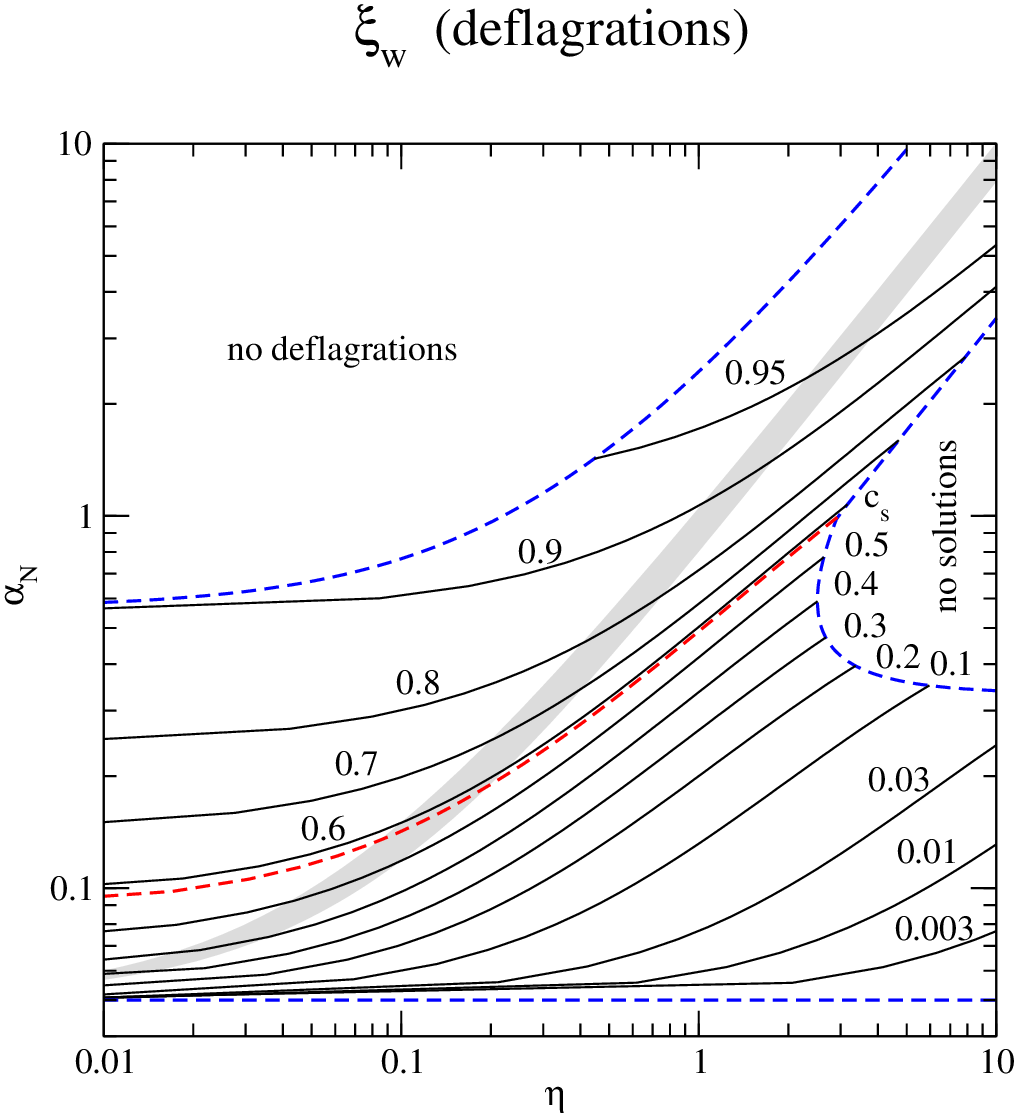}
\caption{\label{fig:conts}
\small Contour plots of $\kappa$ and $\xi_w$ as functions of $\eta$ and
$\alpha_N$ (for $a_-/a_+=0.85$). The blue lines mark the
transition to regions without solutions. The green lines mark the
boundaries between stationary and runaway solutions. The red lines mark
the transition from subsonic  to supersonic deflagrations (hybrids). We 
superimposed the detonation region in the lower plots as a gray band. }
\end{center}
\end{figure}

A fully numerical calculation of $\langle v \rangle$ in a given model 
would require the following procedure. The starting point is the computation of 
the free-energy (in some approximation) as a function of the Higgs field and the 
temperature, ${\cal{F}}(\phi,T)=-p$. First, the nucleation temperature
$T_N$ of the phase transition should be determined (this is standard).
To obtain the steady-state profiles across the phase-transition wall 
in the planar limit of quantities like the velocity, temperature and Higgs field, one should
integrate the following system of coupled differential equations:
\bea
&&\partial^2_z\phi-\frac{\partial 
{\cal{F}}}{\partial\phi}+T_N\tilde{\eta}v\partial_z\phi=0\ ,\nn\\
&&\partial_z[\omega \gamma^2 v]=0\ ,\nn\\
&&\partial_z\left[\frac{1}{2}(\partial_z\phi)^2+\omega \gamma^2 v^2+p
\right]=0\ .
\label{eq:numwall}
\eea
In addition to the first Higgs equation which we already discussed, the two extra
equations correspond to the differential (and static) form of 
energy-momentum conservation. Their integration across the wall gives 
immediately the Steinhardt's matching conditions (\ref{eq:wall_constr}).

One wants to solve the system (\ref{eq:numwall}) for the primary
quantities $\phi(z)$, $T(z)$ and $v(z)$. The boundary conditions for
$\phi(z)$ are 
\be
\phi(-\infty)=\phi_0(T_-) \quad \textrm{and} \quad  \phi(\infty)=0,
\ee
 where $\phi_0(T_-)$ is the Higgs vacuum
expectation value in the broken phase at some temperature $T_-$ to be
determined. The $z=\pm\infty$ boundaries correspond to the assumption that
the width of the phase transition front is much smaller than the width of
the shell with non-zero plasma velocity, which in general is a very good
approximation. For $\tilde\eta$ fixed, the boundary conditions (say at
$z=-\infty$) for $T(z)$ and $v(z)$ cannot be chosen freely: e.g. if one
fixes $T(+\infty)=T_+$ (in general different from $T_N$) only one
particular $v(+\infty)=v_+$ is selected and then all profiles $\phi(z)$,
$T(z)$, $v(z)$ can be determined. Detonation solutions will have
$v(+\infty)=v_+=\xi_w> v(-\infty)=v_-$ and one should choose
$T(+\infty)=T_N$. Deflagrations will have instead $v_+<v_-=\xi_w$ and
$T(+\infty)=T_+$ has to be set consistently with such wall velocity (and
corresponding deflagration hydrodynamic profile).  Finally, hybrid
solutions have $v_+<v_-=c_s^-$. Their wall velocity, $\xi_w>c_s^-$ is not
fixed unequivocally by $v_\pm$ but should match the choice of $T_+$. Our
analytical formula (\ref{eq:micro2}) is expected to give a reasonable
approximation to the more complicated numerical approach just described.  
Such simple formula is very useful to investigate in a model-independent
way the parametric dependences of the wall velocity (and derived
quantities) in different regimes of bubble expansion, which we do next.

Armed with eq.~(\ref{eq:micro2}) we can calculate the wall velocity
for given values of $\alpha_N$ (that measures the strength of the
transition) and $\eta$ (that measures the friction of the plasma). The
results are shown in Fig.~\ref{fig:conts}, which plots contour lines
of the wall velocity $\xi_w$ and the efficiency coefficient $\kappa$
in the plane ($\eta,\alpha_N$), for the particular case of
$a_-/a_+=0.85$ (as in the SM). Let us first explain the different
boundaries of the parameter space available in the plane
$(\eta,\alpha_N)$. For the bubbles to be able to grow, the left hand
side of (\ref{eq:micro2}) should be positive. This requires
$\alpha_+>(1-a_-/a_+)/3$ which, in this particular example, gives
$\alpha_N\geq \alpha_+>0.05$. Both the low horizontal boundary
$(\alpha_N>0.05)$ and the boundary at larger $\alpha_N$ and small
$\eta$ come from this requirement. That bubbles might not be able to
grow even if the transition is in principle quite strong (large
$\alpha_N$) is due to the fact that one can have $\alpha_+\ll\alpha_N$
for hybrid solutions (see figs.~\ref{fig:expAll} and \ref{fig:fit2}).
The boundary at $\alpha_N>1/3$ and large $\eta$ corresponds to the
extreme case with $v_+\rightarrow 0$ and $T_-\rightarrow 0$ and
determines the maximal possible value of the friction coefficient for
a given $\alpha_N$. We have plotted separately the cases of
detonations and deflagrations.  Runaway solutions are realized for
strong transitions and small friction as one would expect. We see that
for weak phase transitions and small friction, deflagrations compete
with detonations (superimposed in the lower plots as a gray band). In
the overlapping region, the wall velocity for the deflagration
solution is always smaller than for the detonation solution, which
could have indicated that the deflagration solution is reached first
and is therefore realized. However, hydrodynamic simulations indicate
that these solutions turn out to be unstable globally and in this
regime the phase transition proceeds by detonation
bubbles~\cite{KurkiSuonio:1995pp} (or eventually by runaway behavior
if the upper bound on friction would be incorporated in the
model). Henceforth, the only way of realizing supersonic deflagrations
is in a regime where hybrids but no detonations are
possible~\cite{KurkiSuonio:1995pp}.

\subsection{Microscopic determination of $\eta$}

In the following we provide the connection of our results to the work
\cite{Moore:1995si,John:2000zq} in which the friction in the wall was
explicitly calculated in the SM and MSSM\footnote{We consider those
results more reliable than the ones of refs.~\cite{Moore:2000wx,
Megevand:2009ut} which deduce particle distribution functions $\delta
f$ without taking interactions into account.}. The friction force in
ref.~\cite{John:2000zq} is of the form
\be
F_{fr} = \hat \eta \left< v \right> T_N 
\int dz \ \phi^2 \ (\partial_z \phi)^2,
\ee
where $\hat \eta$ is a constant that depends mostly on the particle
content of the model and its couplings to the Higgs.
$\phi(z) = \frac12 \phi_N [1 + \tanh (z/l_w)]$ where $l_w$
is the wall thickness and $\phi_N$ is the
Higgs vev at nucleation temperature. Comparison with our friction
force shows that our parameter $\eta$ can be written as
\be
\eta \sim \frac{\hat \eta}{10 \, a_+} \frac1{T_N l_w} \lp \frac{\phi_N}{T_N} \rp^4
\ee
 The coefficient
$\hat \eta$ was determined in the SM \cite{Moore:1995si} ($\hat
\eta \approx 3$) and in the MSSM~\cite{John:2000zq} ($\hat
\eta \lesssim 100$ with a sizable dependence on $\tan \beta$).
A particularly interesting case is given by the parameter region of
the MSSM that allows for viable electroweak baryogenesis. The bound on
sphaleron wash-out implies $\phi_N/T_N \gtrsim 1$ and using $T_N l_w
\approx 10$, $\hat \eta \approx 100$ one finds $\eta \approx 1/30$. Due
to a small difference in free energies, this leads to subsonic wall
velocities $\left< v \right> = 0.05 \div 0.1$~\cite{John:2000zq} as
required for the diffusion of CP-violating particle densities into the
symmetric phase in front of the wall. This corresponds to a very weak
phase transition with a value of $\alpha_N$ just slightly above its
lower bound (that depends on $a_-/a_+$). Note that for models with a
similar particle content the friction $\eta$ is not expected to change
much, while the strength of the phase transition can increase
significantly. This is for example the case in singlet extensions of
the SM and MSSM which can easily lead to detonations or runaway
solutions.

\hskip 0.5 cm

In this section we have assumed that the bubble wall reaches at some (not
too late) stage of the expansion a constant velocity. In this case the
fraction of energy transformed into kinetic energy of the Higgs field
becomes negligible, since it only scales with the surface of the bubble,
while the similarity solutions of bulk motion scale with the volume. This
can change in cases in which the wall keeps accelerating without reaching
a terminal velocity, as discussed in the next section.

\section{Runaway walls\label{sec_runaway}}

It was recently argued~\cite{BM} that the friction exerted on
the Higgs wall by the plasma might be too small and the wall might
continuously accelerate. In this case a constant fraction of the free
vacuum energy is transformed into kinetic and gradient energies of the
wall. In this section we analyze the energy balance and the efficiency
coefficient in this situation.

Let us first quickly present the main result of \cite{BM} that is
based on the analysis of refs.~\cite{Liu:1992tn, Dine:1992wr,
Arnold:1993wc}. The passing phase-transition wall disturbs the
distribution functions of particles in the plasma. As discussed in the
previous section, if we knew such non-equilibrium distributions,
$f_i(p,z)$, for each particle species, we could write, for the total
force acting on the wall per unit area and including friction:
\be
\label{Ftot}
F_{tot}=F_{dr}-F_{fr}=
\Delta V_0+
\sum_i |N_i|\int d z \frac{d m_i^2}{d z}\int 
\frac{d^3p}{(2\pi)^3}\frac{f_i}{2E_i}\ .
\ee
This has the same form as eq.~(\ref{Fdriving}) for the driving force 
$F_{dr}$ but with the replacement $f_i^{eq}\rightarrow f_i$. Now, the 
ultra-relativistic case is particularly simple: to leading order in 
$1/\gamma_w$, the wall induces a sudden 
change in particle masses, $m_{i,+}^2\rightarrow m_{i,-}^2$, but leaves 
particle distribution functions as they were in the symmetric phase
$f_i=f_{i,+}^{eq}$ (which are not the equilibrium ones in the 
broken phase). This allows the $z$-integral in (\ref{Ftot}) to be 
performed and one obtains 
\be
\label{Ftotrel}
F_{tot}=\Delta V_0 -
\sum_i |N_i|\Delta m_i^2\int 
\frac{d^3p}{(2\pi)^3}\frac{f_{i,+}^{eq}}{2E_{i,+}}\ ,
\ee
where $\Delta m_i^2=\left. m_i^2\right|^-_+$ and $E_{i,+}$ in the 
momentum integral is also the one corresponding to the symmetric phase, 
as indicated.

If $F_{tot}$ remains positive even for $v \to 1$, the system will
enter the runaway regime. Besides the explicit form (\ref{Ftotrel}),
this condition can be rephrased in terms of the free energies in the
mean field approximation~\cite{BM}. Notice that the mean field
approximation implies
\be
\left< V_T(\phi) \right>_{MF} = V_T(0)+\sum_i[m_i^2(\phi)-m_i^2(0)]\left.
\frac{d V_T}{d m_i^2}\right|_0,
\ee
such that 
\be
\left< V_T^- \right>_{MF} = \left< V_T^+ \right>_{MF} + 
\sum_i \, \Delta m_i^2  \left.
\frac{d V_T}{d m_i^2}\right|_+,
\ee
and the criterion for runaway behaviour reads
\be
0 < F_{tot} = \left< {\cal F}^+ - {\cal F}^- \right>_{MF}.
\ee

\begin{figure}[ht]
\begin{center}
\includegraphics[width=0.65\textwidth, clip ]{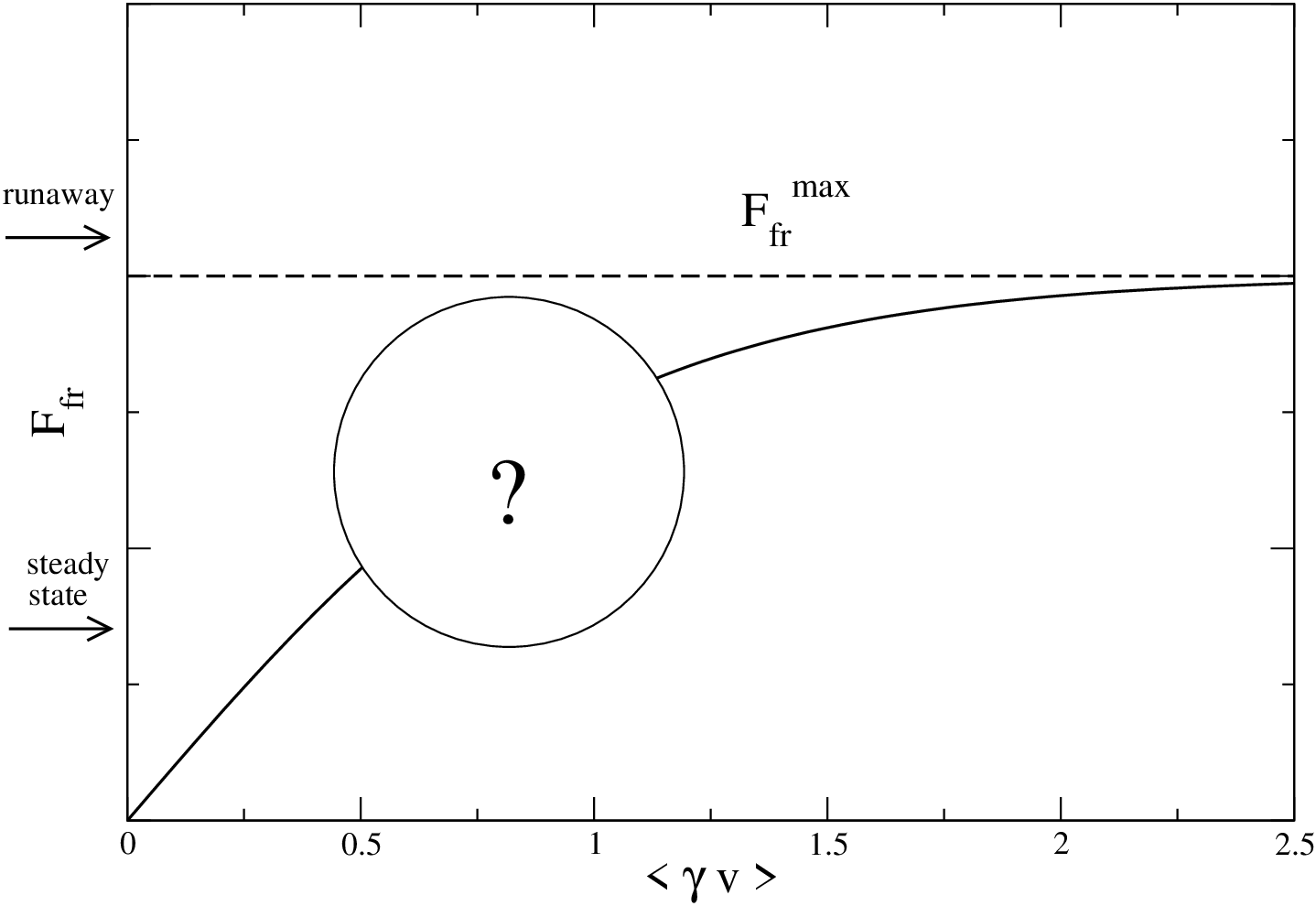}
\caption{\label{fig:friction}
\small 
A sketch of the friction force as a function of the wall velocity
showing the saturation at $\left<v\right>\to1$. The behavior for
intermediate velocities is largely unknown. The arrows indicate two possible values for the 
driving force that would lead to steady or runaway bubble expansion, as indicated.}
\end{center}
\end{figure}

The results we just summarized can be interpreted as saying that the
friction $F_{fr}$ (which we expect to increase monotonically with $v$)
saturates at a finite value\footnote{Up to a possible logarithmic
increase with $\log\gamma_w$ (we thank Guy Moore for clarifications
on this point).} for $v\rightarrow 1$. A sketch for this behavior is
given in Fig.~\ref{fig:friction}.  Noting that the driving force is
given by
\be
 F_{dr} = {\cal F}^- - {\cal F}^+,
\ee
then the maximal value for the friction force reads ($\left< {\cal
F}^+ \right>_{MF} = {\cal F}^+$)
\be
F_{fr}^{max}= \left< {\cal F}^- \right>_{MF} - {\cal F}^- 
= \left< V_T^- \right>_{MF} - V_T^-
\ee
Using the convexity properties of the different contributions to $V_T$
one can show that $F_{fr}^{max}$ is necessarily positive. 

If $\left< {\cal F}^+ - {\cal F}^- \right>_{MF}>0$, then
$F_{dr}<F_{fr}^{max}$, and the wall velocity will grow till the
friction force equilibrates $F_{dr}$ and a steady state with some
terminal velocity is reached. In the opposite case with $\left< {\cal
F}^+ - {\cal F}^- \right>_{MF}<0$, one has $F_{dr}>F_{fr}^{max}$ and
the wall will keep accelerating without reaching a steady state,
i.e. it will ``run-away". The crucial assumptions here are that there
are no hydrodynamic obstacles that prohibit the wall velocities to
become highly relativistic in the first place, and that the mean
free-path of the particles is larger than the wall thickness.

It is instructive to rederive the results of ref.~\cite{BM} we have
just presented in the language of Kadanoff-Baym equations (see
refs.~\cite{Prokopec:2003pj} for an introduction to Kadanoff-Baym
equations in Wigner space). In this formalism, the Wightman function
$G_i^<$ (for particle species $i$) is the relevant Green function that
encodes the distribution function of the particles in the
plasma. Under the assumption that the mean free-path of the particles
that obtain a mass by the Higgs vacuum expectation value is much
larger than the thickness of the bubble wall, one can neglect
collisions in the wall. In this case the Kadanoff-Baym equation for
the Wightman function $G^<_i$ reads
\be
\label{eq:Wightman}
(p^2 - m_i^2 ) \,  e^{i \diamond/2}  \, G^<=0,
\ee
where the diamond operator is defined as
\be
\diamond = \overleftarrow{\partial}_x \overrightarrow{\partial}_p 
- \overleftarrow{\partial}_p \overrightarrow{\partial}_x.
\ee
In the semi-classical limit, the operator $e^{i\diamond/2}$ can be 
expanded in gradients and the real/imaginary parts
of the equations become, at first order,
\bea
(p^2 - m_i^2 ) G_i^< &=& 0, \\
(p^2 - m_i^2 ) \diamond G_i^< &=& 0. 
\eea
The gradient expansion is justified in the present case because in
the wall frame the particles have momenta of order $\gamma T$, which is
large compared to the inverse wall thickness. Using the ansatz
\be
G_i^< = 4\pi \, \delta(p^2 - m_i^2) f_i( x^\mu, p^\mu)\ ,
\ee
this yields for the particle distribution function the equation 
\be
\label{eq:eomf}
\left[ p \cdot \partial_x + \frac12 (\partial_\zeta m_i^2) \lambda \cdot
 \partial_p \right] f_i( x^\mu, p^\mu) = 0. 
\ee
We introduced again a four-vector $\lambda_\mu$ to parametrize the
motion of the wall [that is,  $\lambda_\mu=(0,0,0,1)$ in the frame moving 
with the wall] and $\zeta\equiv \lambda \cdot x$. Notice that the first 
term in (\ref{eq:eomf}) is the flow term of a Boltzmann equation while the 
second term represents
the force from the wall acting on the plasma. In front of the wall the
distribution function is given by the equilibrium one. E.g., for a bosonic 
degree of freedom
\be
f_{i,+}^{eq} = \frac{\theta(p_0)}{\exp [\beta (u \cdot p)] - 1}\ ,
\ee
where $u^\mu$ is the plasma velocity four-vector.
Solving eq.~(\ref{eq:eomf}) the distribution function behind the wall
reads (with $\Delta m_i^2$ as before)
\be
\label{eq:distrfunc}
f_i = \frac{\theta(p_0) }
{\exp \beta \, \Omega_i - 1 }, 
\ee
with
\be
\Omega_i = u \cdot p + u \cdot \lambda 
\left[ \, \lambda \cdot p - 
\textrm{sign}(\lambda \cdot p) \sqrt{ (\lambda \cdot p)^2 + \Delta m_i^2} 
\right].
\ee
In the wall frame this yields
\be
\Omega_i = \gamma p_0 - \gamma v \sqrt{p_z^2 + \Delta m_i^2}\ .
\ee
This is in accord with the results of \cite{BM} and our previous 
discussion: in the wall frame the particles just cross the wall and change 
their momentum according to
\be
\label{shift}
p_z^2 \to p_z^2 - \Delta m_i^2.
\ee
The contribution of particle species $i$ to the pressure behind the wall 
is (in the wall frame and per degree of freedom) given by
\bea
\delta_i T^{plasma}_{zz} &=& \int \frac{d^4p}{(2 \pi)^4} 
\, p_z^2 \, G_i^< \nn \\
&=& 2 \int \frac{d^4p}{(2 \pi)^3} \delta(p^2 - m_{i,-}^2)  
\, p_z^2 \,  f_i \nn \\
&\approx& 2 \int \frac{d^4p}{(2 \pi)^3}  \delta(p^2 - m_{i,+}^2)  
\, p_z \,  \sqrt{p_z^2 - \Delta m_i^2} \, \textrm{sign} \, (p_z)
\ \theta(p^2_z - \Delta m_i^2) \, f_{i,+}^{eq}, 
\eea
where the last line is obtained by shifting the integration variable
as in (\ref{shift}) and noting that the measure $dp_z \, p_z$ does not
change. Hence, for the pressure difference produced by the plasma this
gives
\bea
\delta_i\Delta T_{zz} &\approx&   \Delta m_i^2 \, \int 
\frac{d^4p}{(2 \pi)^3} \delta(p^2 - m_{i,+}^2) f_{i,+}^{eq} \nn \\
&=& \Delta m_i^2 \, \int 
\frac{d^3p}{(2 \pi)^3 \, 2E}  f_{i,+}^{eq},
\eea
which agrees with the formula obtained in ref.~\cite{BM}. In terms of 
$\Delta T_{zz}$, the criterion for runaway solutions is very transparent 
and is simply 
\be
\epsilon>\Delta T_{zz} = \sum_i \,\delta_i\Delta T_{zz} .
\ee
 In this case, only a part of the 
available vacuum energy is released into the plasma and the remaining 
energy is used to further accelerate the wall. 

This criterion for runaway behavior can be reformulated in terms of
parameters of the phase transition in case it is rather strong. For
particles that are light in both phases, the contribution to the
pressure is similar in both phases. Particles that are heavy in both
phases do not contribute much to either pressure difference or free
energy. Hence, mostly the particles that become heavy during the phase
transition produce a pressure difference along the wall. Using this, 
one obtains for an accelerated wall the criterion
\be
\label{eq:BMconstr}
\Delta T_{zz} = \frac{T_N^2}{24} \sum_{light\to heavy} c_i \, |N_i| \, 
m_i^2 
= \frac{T_N^2}{24} \left< \phi \right>^2 \sum_{light\to heavy}
 c_i \, |N_i| \, y_i^2\; < \; \epsilon \ ,
\ee
with $c_i=1$ ($1/2$) for bosons (fermions), $|N_i|$ are the
corresponding numbers of degrees of freedom, $y_i$ are the coupling
strengths to the Higgs boson and $\left< \phi \right>$ the Higgs
vacuum expectation value in the broken phase. Using the relation
$\epsilon= \alpha_N (a_N T_N^4)$ we get that a runaway wall is in
principle possible for
\be
\label{eq:alpha_bound}
\alpha_N > \alpha_\infty \equiv 
\frac{30}{\pi^2}   \left(\frac{\left< \phi \right>}{T_N}\right)^2 
\frac{\sum_{light\to heavy} c_i \, |N_i| \, y_i^2}
{\sum_{light} c'_i\, |N_i|}\ ,
\ee
with $c'_i=1$ ($7/8$) for bosons (fermions). This equation serves as
the definition of $\alpha_\infty$.  In extensions of the SM,
eventually more particles contribute to the pressure difference, but
typically not many new light particles are in thermal equilibrium. One
hence can deduce that for
\be
\alpha_N > 1.5 \times 10^{-2} 
\left(\frac{\left< \phi \right>}{T_N}\right)^2 ,
\ee
runaway walls are possible depending on the details of the model. It
is interesting to note that models which lead to sizable gravitational
wave production typically satisfy the runaway condition and this
should be taken into account when calculating the GW signal.

Next, we make contact with the case of a constant wall velocity,
discussed in the last section.  

\section{Energy budget of first-order phase transitions}
\label{sec:summary}

The analysis in the last section assumed that the system was
time-independent in the wall frame, which leads to the fact that the
Higgs field only contributes a pressure component from the vacuum
energy to the energy momentum tensor
\be
 T^\phi_{zz}\left. \right|^+_- = - \epsilon\ , \quad
 T^\phi_{0z}\left. \right|^+_- = 0\ .
\ee
In a static system these contributions have to be compensated by the
plasma and this requires
\be
\label{eq:constr2}
 T^{plasma}_{zz} \left.\right|^+_- =  \epsilon\ , \quad
 T^{plasma}_{0z} \left.\right|^+_- = 0\ .
\ee
Such relations lead to the matching conditions (\ref{eq:wall_constr})
used as boundary conditions in the hydrodynamic analysis of the
plasma. In the case of a highly relativistic plasma, these boundary
conditions can be derived explicitly from the particle distribution
functions. Following the same calculation as in the previous section,
for $T^{plasma}_{0z}$ one finds
\be
\label{eq:constr3}
  T^{plasma}_{zz}\left. \right|^+_- = \Delta T_{zz}, \quad
  T^{plasma}_{0z}\left. \right|^+_- = 0,
\ee
Hence, in the runaway case, with $\alpha_N>\alpha_\infty$ the
solutions for the fluid motion are identical to the ones with
$\alpha_N=\alpha_\infty$, according to the distribution functions
determined close to the wall. At the same time the Higgs field cannot
be time-independent anymore and energy momentum conservation implies
that the remaining energy is used to accelerate the wall.

\begin{figure}[t]
\begin{center}
\includegraphics[width=0.90\textwidth, clip ]{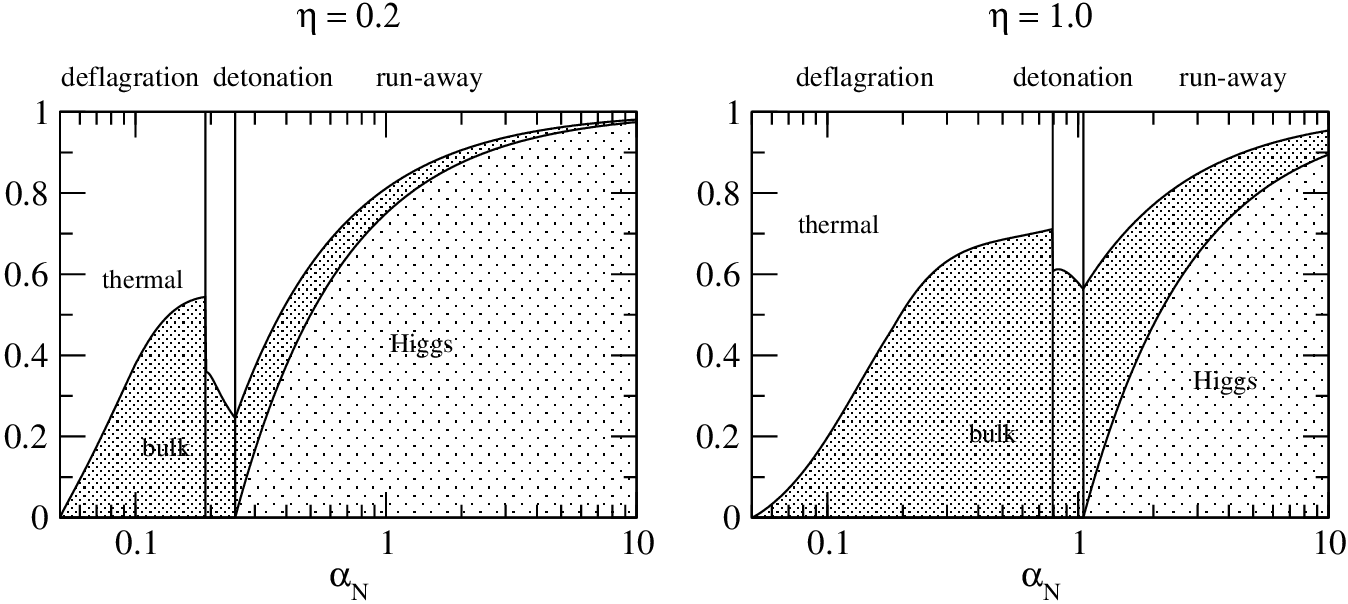}
\caption{\label{fig:budget}
\small The energy budget for $\eta=0.2$ and $\eta=1.0$. The different
contributions (from top to bottom) are thermal energy, bulk fluid
motion and energy in the Higgs field. The last two components can
potentially produce anisotropic stress in the plasma and subsequently
gravity waves.}
\end{center}
\end{figure}

We observed in section 4 that, in the limit of large wall velocities, the 
efficiency
factor does not depend on the wall velocity but is given by
(\ref{eq:kapBMstat}). This means that, in the runaway case,
\be
\label{eq:kapBMrunaway}
\kappa_\infty \simeq 
\frac{\alpha_\infty}{0.73 + 0.083 \sqrt{\alpha_\infty} +  \alpha_\infty}
\quad \textrm{(runaway)}.
\ee
In summary, in the runaway regime and for given $\alpha_N$, a portion
$\alpha_\infty$ of the initial $\alpha_N$ produces bulk motion with
efficiency $\kappa_\infty$, as given by eq.~(\ref{eq:kapBMrunaway}),
while the remaining portion, $\alpha_N-\alpha_\infty$, is transformed
directly into kinetic/gradient energy of the Higgs field with
efficiency $\kappa=1$. These two components can potentially produce
anisotropic stress in the plasma and subsequently gravity waves while
the thermal energy in the plasma can not.  Figure~\ref{fig:budget}
shows the energy budget of the phase transition for two choices of the
friction coefficient $\eta$ as a function of $\alpha_N$ in different
regimes of bubble expansion.

\section{Summary}
\label{sec:conclusion}

The bubble wall velocity $\xi_w$ in first-order phase transitions is a
key quantity entering the calculation of the baryon asymmetry in
electroweak baryogenesis and its derivation has been discussed
extensively in the literature. However, it has been treated in detail
only in specific models (corresponding to weak first-order phase
transitions) and a general account of the problem was lacking. In this
work, we attempted to gather all the important information in a
self-consistent manner and in a model-independent approach.  We presented
a unified description of all the different regimes characterizing
bubble growth and stressed how they are connected.

In the last few years, there has been a significant effort towards
working out the relic gravity wave background generated by bubble
collisions. Production of gravity waves during a first-order phase
transition is due to bulk motions in the plasma and magnetic fields
generated by these fluid motions. The resulting gravity wave spectrum
scales as a large power of the typical fluid velocity, roughly as
$v^4$. The fluid velocity profile in the vicinity of the bubble wall
is relevant and it is therefore important to estimate the fraction
$\kappa$ of the vacuum energy density $\epsilon$ liberated during the
phase transition which goes into the fluid motions.
 
So, $\xi_w$ and $\kappa$ are the two salient quantities that one would
like to predict for any given particle physics model leading to a
first-order phase transition.  The most straightforward quantity to
compute in a given model is $\alpha_N$, the ratio of the vacuum energy
density $\epsilon$ to the radiation energy density at the nucleation
temperature. One would be interested in knowing in which regime
(detonation, deflagration, hybrids, runaway) the given model is
expected to fall in, without having to deal with the intricate
Boltzmann equations.  There is a relation which has been used
extensively in the literature for $\xi_w$ as a function of
$\alpha_N$. Such relation is valid for supersonic walls and is based
on the assumption of Jouguet detonations. Assuming Jouguet detonations
amounts to setting the value of the fluid velocity at the inner
boundary of the velocity profile in the wall frame to the speed of
sound. There is no justification for this choice although it has been
used in the literature for simplicity.  Setting arbitrarily some
boundary condition for the velocity leads to inconsistencies and
corresponds to ignoring the constraints from the equation of motion of
the Higgs field. By dealing explicitly with that Higgs equation,  
our study goes beyond this assumption.

The difficulty is that the problem under consideration is not fully determined 
by $\alpha_N$.  As we elucidated, for a
given $\alpha_N$, there are many possible wall velocities and a large
range of $\kappa$ values\footnote{We have also corrected a missing factor $1/\xi_w^3$ in
the definition of $\kappa$ which is found in the literature.}, as seen in Fig.~\ref{fig:KfuncH}.  To fix the
solution, one needs to compute the friction term that restrains the
bubble expansion. There are two approaches to this problem.  The
rigorous (but tedious) treatment requires solving coupled Boltzmann
equations.  A simpler approach is to take the friction as a
parameter independent from the bubble wall velocity and solve the equation of
motion for the Higgs field by using some phenomenological description
for the friction.  This is the approach that we took in this paper.
It has been adopted in some numerical studies~\cite{Laine:1993ey,
Ignatius:1993qn, KurkiSuonio:1984ba, Gyulassy:1983rq,
KurkiSuonio:1995pp}, although the modeling of the friction was
not appropriate  in the limit of large velocities. We now use a more
realistic modeling which allows us to describe the runaway regime in
addition to the steady state regime, eq.~(\ref{HEOMw}).

One of our final results has been to present, in Fig.~\ref{fig:conts},
model-independent contours for $\xi_w$ and $\kappa$ in the
$\eta-\alpha_N$ plane, where $\eta $ and $\alpha_N$ are the two crucial
physical (dimensionless) parameters respectively characterizing the
strength of the phase transition and the amount of friction. 
%(so far no
%description as a function of $\eta$ had been proposed in the
%literature; besides for deflagrations there was no relation at all
%available for the wall velocity $\xi_w$ and efficiency $\kappa$ in
%terms of $\alpha_N$). 
 In concrete models, $\alpha_N$ and $\eta$ are not
independent parameters. However, 
%the actual calculation of these quantities is quite involved and 
we believe that the information
contained in these plots is useful as it
enables to investigate the parametric dependence of the wall velocity
and the $\kappa$ factor in different regimes and gather all the
important physics in a single comprehensive figure. The contour plots are not fully
model-independent as we have fixed the relative
change in the number of relativistic degrees of freedom between the
two phases but they serve the purpose of describing the qualitative
behavior of $\xi_w$ and $\kappa$.  From our analysis, we obtain that
only a small band in the $(\eta,\alpha_N)$ plane leads to pure
detonations.  We also find that in a large region of this plane,
hybrids coexist with runaway solutions, meaning that our
time-independent analysis is not sufficient to fix the solution.
According to numerical studies, hybrids are mostly
unstable~\cite{KurkiSuonio:1995pp}. We would therefore conclude that
the most likely solutions are either deflagrations or runaway
solutions. We have clarified the condition determining the runaway regime, see
eq.~(\ref{eq:alpha_bound}), and found that for values $\alpha_N \sim
{\cal O}(1)$ common in the literature, a runaway
regime is quite likely.
Note however than numerical studies have not been carried
out in the ultra relativistic regime.

 For deflagrations and hybrids, a significant
fraction of the available energy goes into bulk fluid motions. On the
other hand, for a non-steady state solution, a large fraction of the
energy goes into accelerating the wall and little goes into bulk fluid
motions, as seen in Fig.~\ref{fig:budget}. Therefore, and perhaps
counterintuitively,  we find that, in a very strong first order phase transition,
the contribution to the gravity wave spectrum from turbulent fluid
flows is probably subdominant. While the overall size of the gravity
wave signal is essentially controlled by the amount of vacuum energy
released, i.e. by  $\alpha_N$, its detailed spectrum will depend on how the
energy is distributed among the different components (wall versus
plasma).  We have therefore provided the relation between the fluid
velocity and the bubble wall velocity in Fig.~\ref{fig:flow_max}. In
addition, we estimated the thickness of the plasma shell near the
bubble wall where the kinetic energy in the plasma is concentrated. In
all numerical calculations of the GW background
\cite{Kosowsky:1991ua,Kamionkowski:1993fg,Huber:2008hg}, it has been
assumed for simplicity that all the energy was concentrated on the
bubble surface and finite thickness effects were ignored (they were
instead considered in the analytical approach of
\cite{Caprini:2007xq}). This is a reasonable approximation for
detonations, although we show that even for supersonic wall
velocities, the thickness of the plasma shell can reach $\sim 20 \%$
of the bubble size, see Fig.~\ref{fig:thick}.

Finally, the nature of the electroweak phase transition is unknown and
it will take some time before we can determine whether electroweak
symmetry breaking is purely Standard Model-like or there are large
deviations in the Higgs sector which could have led to a first-order
phase transition.  Although our analysis was implicitly motivated by the
electroweak phase transition, it could be applicable  to other
 first-order phase transitions in the early universe.

\section*{Acknowledgment}
We thank Dan Chung for mentioning the mismatch in the different
definitions of efficiency factors to us. J.R.E. and J.M. No thank CERN
TH-Division for partial financial support during different stages of
this work.  We acknowledge support from the European Commission the
European Research Council Starting Grant Cosmo@LHC and the Marie Curie
Research and Training Networks ``ForcesUniverse" (MRTN-CT-2006-035863)
and 'UniverseNet' (MRTN-CT-2006-035863); by the Spanish
Consolider-Ingenio 2010 Programme CPAN (CSD2007-00042); the Comunidad
Aut\'onoma de Madrid under grant HEPHACOS P-ESP-00346 and the Spanish
Ministry MICNN under contract FPA 2007-60252.

\appendix

\section{Numerical fits to the efficiency coefficients\label{sec_fits}}

In this section we provide fits to the numerical results of
section~\ref{sec_eff}. These fits facilitate the functions
$\kappa(\xi_w, \alpha_N)$ and $\alpha_+(\xi_w, \alpha_N)$ without
solving the flow equations and with a precision better that $15\%$ in
the region $10^{-3} < \alpha_N < 10$.

In order to fit the function $\kappa(\xi_w, \alpha_N)$, we split the
parameter space into three regions and provide approximations for the
four boundary cases and three families of functions that interpolate
in-between: For small wall velocities one obtains ($\xi_w \ll c_s$)
\be
\kappa_A \simeq \xi_w^{6/5} 
\frac{6.9 \alpha_N}{1.36 - 0.037 \sqrt{\alpha_N} + \alpha_N}\ .
\ee
For the transition from subsonic to supersonic deflagrations ($\xi_w =
c_s$)
\be
\kappa_B \simeq \frac{\alpha_N^{2/5}}{0.017+ (0.997 + 
\alpha_N)^{2/5} }\ .
\ee
For Jouguet detonations ($\xi_w = \xi_J$), as stated in
eq.~(\ref{eq:JDetKappa})
\be
\kappa_C \simeq \frac{\sqrt{\alpha_N}}
{0.135 + \sqrt{0.98 + \alpha_N}}\quad \textrm{ and } \quad
\xi_J = \frac{\sqrt{ \frac23 \alpha_N  + \alpha_N^2 } + \sqrt{1/3}}
{1 + \alpha_N}\ .
\ee
And finally for very large wall velocity, ($ \xi_w \to 1$) as stated in
eq.~(\ref{eq:kapBMstat})
\be
\kappa_D \simeq \frac{\alpha_N}
{0.73 + 0.083 \sqrt{\alpha_N} + \alpha_N}\ .
\ee

\begin{figure}[t]
\begin{center}
\includegraphics[width=0.65\textwidth, clip ]{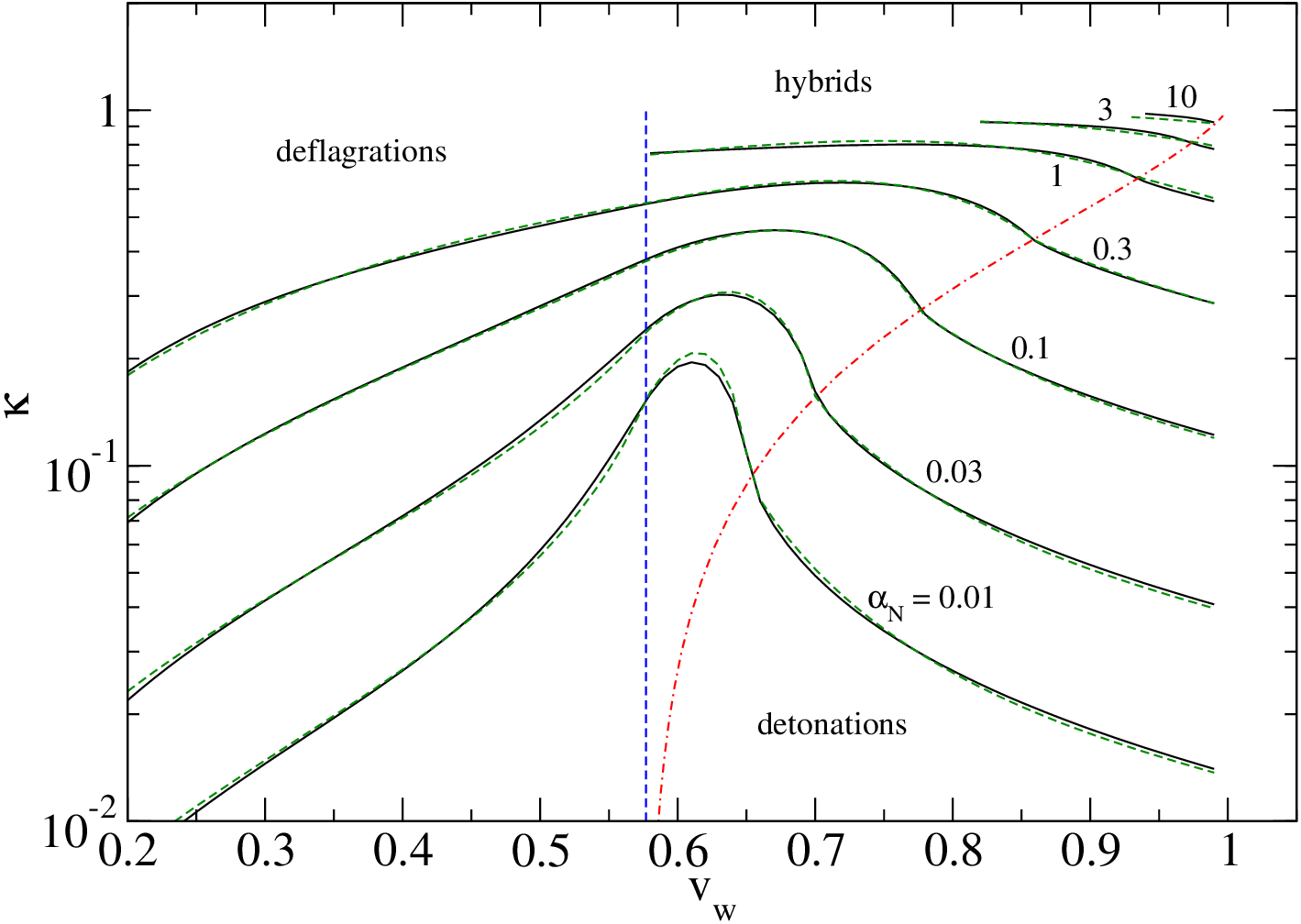}
\caption{\label{fig:fit1}
\small The efficiency coefficient $\kappa$ as a function of the wall velocity
$\xi_w$ for fixed $\alpha_N$ and the fit (dashed lines) described in 
the text. }
\end{center}
\end{figure}

For subsonic deflagrations a good fit to the numerical results is
provided by
\be
\kappa( \xi_w \lesssim c_s) \simeq 
\frac{ c_s^{11/5}
\kappa_A \kappa_B }
{(c_s^{11/5} -  \xi_w^{11/5} )\kappa_B
+  \xi_w c_s^{6/5} \kappa_A}\ ,
\ee
and for detonations by
\be
\kappa( \xi_w \gtrsim \xi_J) \simeq
\frac{ (\xi_J - 1)^3 \xi_J^{5/2}  \xi_w^{-5/2}
\kappa_C \kappa_D }
{[( \xi_J -1)^3 - ( \xi_w-1)^3] \xi_J^{5/2} \kappa_C
+ ( \xi_w - 1)^3 \kappa_D }\ .
\ee
The numerical result for the hybrid (supersonic deflagration) region
is well described by a cubic polynomial. As boundary conditions, one
best uses the two values of $\kappa$ and the first derivative of
$\kappa$ at $ \xi_w=c_s$. Notice that the derivative of $\kappa$ in $
\xi_w$ is not continuous at the point $\xi_J$. The derivative
at $ \xi_w=c_s$ is approximately given by
\be
\delta\kappa \simeq -0.9 \log{\frac{\sqrt{\alpha_N}}
{1 + \sqrt{\alpha_N}}}\ .
\ee
This differs from the derivative one would obtain from the fit in the
region $ \xi_w<c_s$, but mostly for values $\alpha \gtrsim 1$, where
no solutions exist for $ \xi_w<c_s$. The expression for supersonic
deflagrations then reads
\be
\kappa(c_s <  \xi_w < \xi_J ) \simeq \kappa_B + 
( \xi_w - c_s) \delta\kappa + \frac{( \xi_w - c_s)^3}{ (\xi_J - c_s)^3} 
[ \kappa_C - \kappa_B -(\xi_J - c_s) \delta\kappa ]\ .
\ee
The fits of $\kappa$ compared to the numerical results are given in
Fig.~\ref{fig:fit1}. The relative errors never exceed $15\%$.

\begin{figure}[ht]
\begin{center}
\includegraphics[width=0.65\textwidth, clip ]{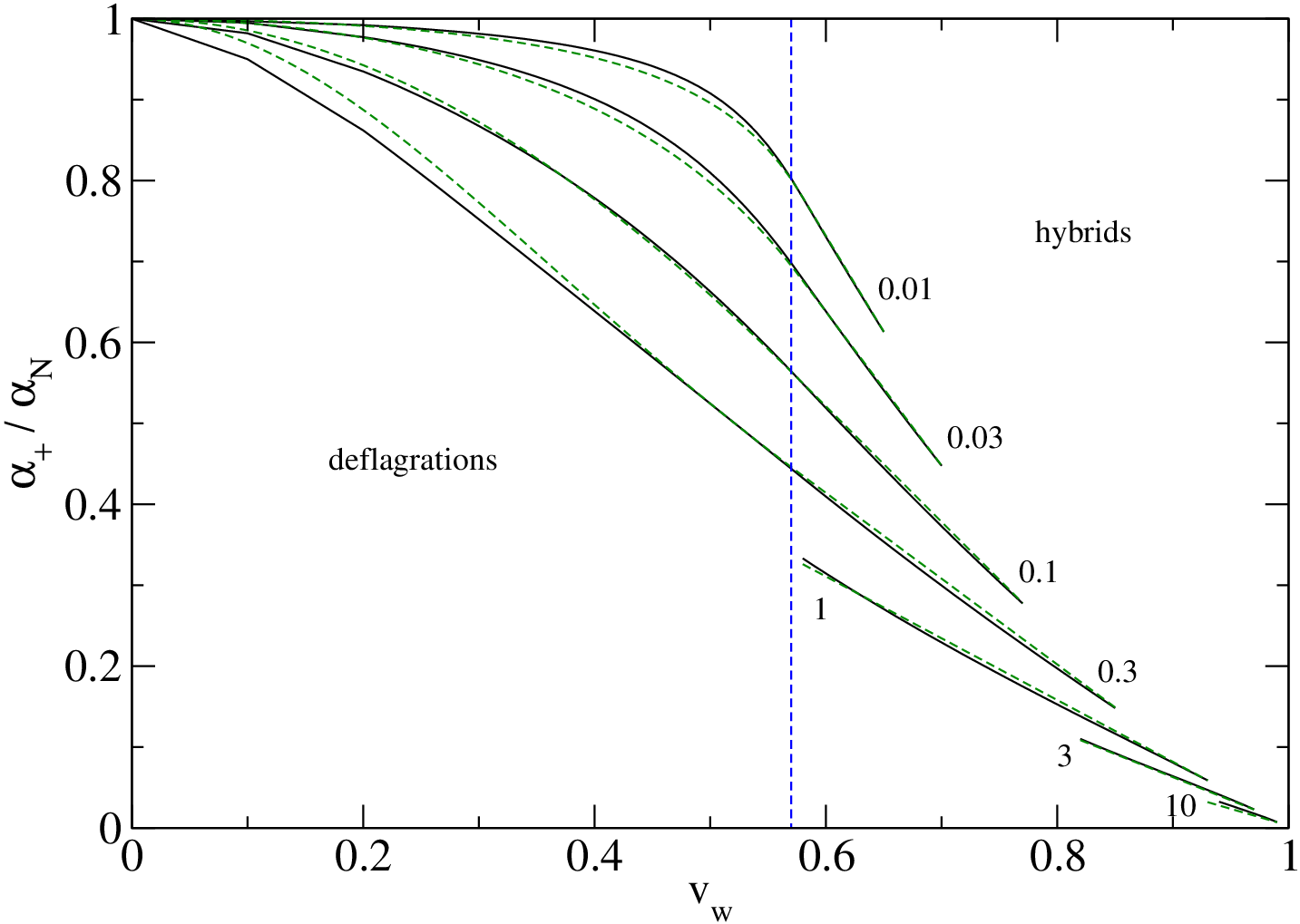}
\caption{\label{fig:fit2}
\small The ratio $\alpha_+ / \alpha_N$ as a function of the wall velocity
$\xi_w$ for fixed $\alpha_N$ and the fit (dashed lines) described in the 
text.  }
\end{center}
\end{figure}

Another useful function is the maximal $\alpha_N$ that can be realized
for fixed $ \xi_w$. This function is approximately given by
\be
\label{eq:alpha_max}
\alpha_N^{max} \simeq \frac13 (1 -  \xi_w)^{-13/10}\ . 
\ee
Finally, the connection between $\alpha_N$ and $\alpha_+$ might be
needed in phenomenological studies. Like in the case for $\kappa$, we
provide approximations or analytic expressions in limiting cases and
interpolate in between. For velocities larger than the Jouguet velocity
$\xi_J$, $\alpha_+$ and $\alpha_N$ coincide. For $\alpha_N<1/3$ all
velocities can be realized and in this case the limit $v_w \to 0$ also
yields $\alpha_+ \to \alpha_N$. Otherwise there is a minimal $ \xi_w$
which is, according to eq.~(\ref{eq:alpha_max}), given by 
\be
\alpha_+= 1/3
\quad \textrm{ for } \quad
 \xi_w = 1 - (3 \alpha_N)^{-10/13}
\quad \textrm{ if } \quad 
\alpha_N > \frac13.
\ee
For relatively small $\alpha_N$ a wall velocity of the speed of sound
can be realized leading to a fit
\bea
\alpha_+ (c_s)&\simeq& \alpha_N 
(0.329- 0.0793 \log{\alpha_N} 
+ 0.0116 \log^2{\alpha_N} +  0.00159 \log^3{\alpha_N})
\nn \\ &&
\quad \textrm{ if } \quad 
\alpha_N < \frac13 (1-c_s)^{-13/10} \simeq 1.02 .
\eea
Finally, the transition from supersonic deflagrations to detonations
can be determined analytically leading to
\be
\alpha_+(\xi_J)= \alpha_N \frac{3 - 3 \xi_J^2}
{9 \xi_J^2 -1}
\quad \textrm{ for all } \quad 
\alpha_N.
\ee
For $\alpha_N>1/3$, the function $\alpha_+ (\alpha_N, \xi_w)$ is well
approximated by a polynomial in $ \xi_w$ that contains the two/three
data points just given for the case $\alpha_N \gtrless 1.02$. For
$\alpha_N<1/3$, the function $\alpha_+ (\alpha_N, \xi_w)$ is
approximately linear in the region $ \xi_w \in [c_s, \xi_J]$ and goes
slowly to $\alpha_N$ for $ \xi_w \to 0$. For velocities below $c_s$, a
reasonable approximation is of the form
\be
\alpha_+ = \frac{\alpha_N + c_1 \,  \xi_w^2}{1 + c_2 \,  \xi_w^2}\ ,
\ee
and matching of this function and its derivative to the linear regime
yields
\be
\alpha_+ = \frac{(c_s^2-\xi_w^2)[\alpha_+(\xi_J)- \alpha_+(c_s) 
]\alpha_N - 2\xi_w^2(1  - \xi_J/c_s) [\alpha_+(c_s) 
-\alpha_N]\alpha_+(c_s) }
{(c_s^2-\xi_w^2)[\alpha_+(\xi_J)- \alpha_+(c_s) ]- 
 2\xi_w^2(1  - \xi_J/c_s ) [\alpha_+(c_s) -\alpha_N] }.
\ee
The comparison between fit and numerical results is given in
Fig.~\ref{fig:fit2}. The relative error never exceeds $5\%$ and is
worst for $\alpha_N \sim 0.3$.

\end{document}